\documentclass[11pt]{article}
\usepackage{amsmath}
\usepackage{amssymb}

\def\beqar {\begin{eqnarray}}
\def\eeqar {\end{eqnarray}}
\def\beq {\begin{equation}}
\def\eeq {\end{equation}}
\def\ra {{\rangle}}
\def\la {{\langle}}
\def\half {{\textstyle{1\over 2}}}
\def\Tr {{\rm Tr}}
\def\tr {{\rm tr}}

\def\del {{\partial}}

\def\ep {{\epsilon}}
\def\vf {{\varphi}}

\def\a {{\alpha}}

\def\bu {{\bar u}}

\def\bz {\bar{z}}
\def\bZ {\bar{Z}}

\def \A {{\mathcal A}}
\def \D {{\mathcal D}}

\def \L {{\mathcal L}}

\def\no2 {{\textstyle{n\over 2}}}

\def \CMP {{ Commun. Math. Phys.}}
\def \PRL {{ Phys. Rev. Lett.}}
\def \PL {{Phys. Lett. B}}
\def \NPBProc {{ Nucl. Phys. B (Proc. Suppl.)}}
\def \NP {{ Nucl. Phys.}}
\def \RMP {{ Rev. Mod. Phys.}}
\def \JGP {{ J. Geom. Phys.}}
\def \CQG {{ Class. Quant. Grav.}}
\def \MPL {{Mod. Phys. Lett.}}
\def \IJMP {{ Int. J. Mod. Phys.}}
\def \JHEP {{ JHEP}}
\def \PR {{Phys. Rev.}}
\def \JMP {{ J. Math. Phys.}}

\begin{document}

\rightline{IC/2004/37}
\rightline{CCNY-HEP-04/5}
\vspace{1cm}
\begin{center}
{\Large \bf Fuzzy spaces, the M(atrix) model and the quantum Hall effect}
\footnote{To appear in the Ian Kogan Memorial volume, {\it
From Fields  to Strings: Circumnavigating Theoretical Physics.}}

\vspace{1.5cm}

DIMITRA~KARABALI\\
Department of Physics and Astronomy\\
Lehman College of the CUNY\\
Bronx, NY 10468\\
E-mail: karabali@lehman.cuny.edu\\
\vspace{0.7cm}

V.~P.~NAIR\\
Physics Department\\
City College of the CUNY\\
New York, NY 10031\\
E-mail: vpn@sci.ccny.cuny.edu\\ 
\vspace{0.7cm}

S.~RANDJBAR-DAEMI\\
Abdus Salam International Centre for Theoretical Physics\\
Trieste, Italy\\
E-mail: seif@ictp.trieste.it
\end{center}
\vskip 0.8cm
\centerline{\bf Abstract}
\bigskip

This is a short review of recent work on fuzzy spaces in their
relation to the M(atrix) theory and the quantum Hall effect. We give an
introduction to fuzzy spaces and how the limit of large matrices is
obtained. The complex projective spaces ${\bf CP}^k$, and to a lesser  
extent
spheres, are considered. Quantum Hall effect and the behavior of edge
excitations
of a droplet of fermions on these spaces and their relation to fuzzy  
spaces
are also discussed.

\newpage
\setcounter{equation}{0}

\section{Introduction}
It is a well known fact that in many quantum mechanical systems, as the
occupation number becomes very large, the
quantum theory can be approximated by a classical theory.
Quantum observables which are linear hermitian operators on the Hilbert  
space
can be approximated by functions on the classical phase space.  
Properties of
classical functions on the phase space can thus be obtained as a limit  
of
the quantum theory.
This raises the possibility that one may consider the quantum Hilbert  
space
and the algebra of
operators on it as the fundamental entities for constructing a manifold,
the classical (phase) space being
only an approximation to it. Fuzzy spaces are a realization of this
possibility \cite{connes1, madore1, landi1, gaw}.
They are defined by a sequence of triples,
$({\mathcal H}_N , Mat_N , \Delta_N )$,
where $Mat_N$ is the matrix algebra of $N\times N$ matrices which act  
on the
$N$-dimensional Hilbert space ${\mathcal H}_N$, and $\Delta_N$ is a  
matrix
analog of the Laplacian.
The inner product on the matrix algebra is given by
$\la A,B\ra = {1\over N} \Tr (A^\dagger B)$.
Such fuzzy spaces may be considered as a finite-state approximation
to a smooth manifold $M$, which will be the classical phase space
corresponding to ${\mathcal H}_N$ as $N\rightarrow \infty$. More  
specifically,
the matrix algebra $Mat_N$ approximates to the algebra of functions on
a smooth manifold $M$ as $N\rightarrow\infty$. The Laplacian
$\Delta_N$ is needed to recover metrical and other geometrical
properties of the manifold
$M$. For example, information about the dimension of $M$ is contained in
the growth of the number of eigenvalues.

Fuzzy spaces are part of the more general framework of
noncommutative geometry of A. Connes and others \cite{connes1, gaw,  
connes}.
Noncommutative geometry is a generalization of ordinary geometry,
motivated by the following observation.
Consider the algebra of complex-valued square-integrable
functions on a manifold $M$. The algebra of such functions with  
pointwise
multiplication is a commutative $C^*$-algebra. It captures many of the
geometrical features of the manifold $M$. Conversely, any commutative
$C^*$-algebra can be represented by the algebra of functions on an
appropriate space $M$. This leads to the idea that a noncommutative
$C^*$-algebra may be considered as the analog of an ``algebra of  
functions''
on some noncommutative space. One can then develop properties of this
noncommutative space in terms of the properties of the algebra. This is  
the
basic idea.

More specifically, one introduces the notion of a spectral triple
$(\A, {\mathcal H}, \D )$, where $\A$ is a noncommutative algebra with  
an
involution,
${\mathcal H}$ is a Hilbert space on which we can realize the algebra  
$\A$
as bounded operators and
$\D$ is a special operator which will characterize the geometry.
In terms of such a spectral triple,
there is a construction of the analog of differential calculus
on a manifold. In particular, if
${\mathcal H}$ is the space of square-integrable spinor functions on a
manifold $M$
(technically, sections of the irreducible spinor bundle), $\A$ is the
algebra of complex-valued
smooth functions on $M$, and $\D$ is the Dirac operator
on $M$ for a particular metric and the Levi-Civita spin connection, then
the usual differential calculus on $M$ can be recovered from the  
spectral
triple.
(For further developments in physics along these lines, see  
\cite{others}.)
In what follows, we shall be interested in fuzzy spaces where we use
matrix algebras to approximate the algebra of functions on a manifold.

A number of fuzzy spaces have been constructed by now.
A finite dimensional Hilbert space is obtained if one quantizes a  
classical
phase
space of finite volume. Thus, for manifolds $M$ which have a symplectic
structure,
so that they can be considered as classical phase spaces, and have  
finite
volume,
we have a natural
method of constructing fuzzy approximations to
$M$. We quantize the phase space $M$ and consider the algebra of  
matrices
acting on the resultant Hilbert space.

A natural family of symplectic manifolds of finite volume are given by
the co-adjoint orbits of a compact semisimple Lie group $G$. (In this  
case,
there is no
real distinction between co-adjoint and adjoint orbits.
For quantization of co-adjoint orbits, see \cite{geomquant, perel}.)
One can quantize such spaces, at least when a Dirac-type quantization
condition is satisfied,
and the resulting Hilbert space corresponds to a unitary irreducible
representation of the group
$G$.  In this way, we can construct fuzzy analogs of
spaces which are the co-adjoint orbits.
In the following, we will work through this strategy for the case of
${\bf CP}^k = SU(k+1) / U(k)$.

In this review, we will focus on fuzzy spaces, how they may appear as
solutions to M(atrix) theory and their connection to generalizations
of the quantum Hall effect. There is a considerable amount of  
interesting work
on noncommutative spaces, particularly flat spaces, in which case one  
has infinite-dimensional
matrices, and the properties
of field theories on them. Such spaces can also arise in special limits
of string theory. We will not discuss them here, since there are  
excellent
reviews on
the subject \cite{douglas}.

\setcounter{equation}{0}

\section{Quantizing ${\bf CP}^k$}

\subsection{The action and the Hilbert space}

We start with some observations on
${\bf CP}^k$.\,\footnote{\,Much of the material in this section
is well known and can be found in  many places; we will follow the  
presentation
in \cite{KN1, KN2}. For an earlier work on coherent states on ${\bf  
CP}^k$
and related matters see \cite{Randjbar-Daemi:1992zj}.} This is the  
complex
projective space of
complex dimension $k$ and  is given by a set of complex
numbers $u_\alpha$, $\alpha = 1, 2, ..., (k+1)$,
with the identification
$u_\alpha \sim \lambda u_\alpha$ for any nonzero complex number
$\lambda$.
We introduce a differential one-form given by
\beq
A(u) = - {i\over 2}~ \left[ { \bu\cdot du - d\bu\cdot u \over
\bu\cdot u}\right]
  \label{1.1}
\eeq
where $\bu \cdot d u = \bu^\alpha du_\alpha$, etc.
Notice that this form is not
invariant under
$u \rightarrow \lambda u$; in fact,
\vspace{-3mm}
\beq
A(\lambda u ) = A(u) ~+~ d f\label{1.2}
\eeq
where $f = -{i\over 2} \log (\lambda /{\bar \lambda})$.
($\lambda$ can, in general, be a function of the coordinates, $\bu$,  
$u$.)
This transformation law shows that the exterior derivative or
curl of $A$ is in fact invariant under  $u \rightarrow \lambda u$;
it is the K\"ahler two-form
of ${\bf CP}^k$ and is given by

The field strength corresponding to this potential is
\beqar
\Omega &\equiv&  dA \nonumber\\[1mm]
&=& -i \left[ {d\bu \cdot du \over \bu\cdot u} -
{d\bu\cdot u ~\bu \cdot du \over (\bu \cdot u)^2}\right]\,.
\label{1.3}
\eeqar
Notice that $\Omega$ is closed, $d \Omega =0$, but it is not exact,
since the form $A$ is not well-defined on the manifold. (We may say that
  $\Omega$
is an element of the second cohomology group of ${\bf CP}^k$.)
The symplectic form we choose to quantize ${\bf CP}^k$ will be  
proportional to
$\Omega$.

The identification of $u$ and $\lambda u$ shows that,
by choosing $\lambda$ appropriately, we may take
$u_\alpha$ to be normalized so that ${\bar u}^\alpha u_\alpha =1$.
In this case, we can introduce local complex coordinates for the
manifold by writing
\beq
u_\a = {1\over \sqrt {1+\bz \cdot z}} \left(
\begin{matrix}
1 \\ z_1\\ .... \\ z_k
\end{matrix}
\right)\ .
\label{1.4}
\eeq
In the local coordinates $z , \bz$, the two-form $\Omega$ has the form
\beq
\Omega=-i\left[ {d\bz_i~ dz_i \over (1+\bz \cdot z)} -{d\bz \cdot z
~\bz\cdot dz
\over (1+\bz \cdot z)^2}\right]\,. \label{1.5}
\eeq

In terms of the normalized $u$'s, a basis for functions on ${\bf CP}^k$  
is
then given by $\{ \phi_l\}$, where $l$ can take all integral values from
zero to infinity,
and
\beq
\phi_l =  \bu^{\alpha_1} \cdots \bu^{\alpha_l} u_{\beta_1}\cdots
u_{\beta_l}\ .\label{1.5a}
\eeq
Notice that, for a fixed value of $l$, we have complete symmetry for all
the upper indices
corresponding to the $\bu$'s and complete symmetry corresponding to the  
lower
indices; further any contraction of indices corresponds to a lower value
of $l$ since $\bu \cdot u =1$. Thus
the number of independent functions is given by
\vspace{-3mm}
\beq
d (k,l) = \left[{ (k+l)! \over k! l! }\right]^2 - \left[{ (k+l-1)!  
\over k!
(l-1)! }\right]^2 .
\label{1.6}
\eeq
Since the traceless part of $\bu^\alpha u_\beta$ transforms as the  
adjoint
representation
of $SU(k+1)$, we see that these functions are contained in the
representations obtained by
reduction of the products of the adjoint with itself.

As we have mentioned before, ${\bf CP}^k$ can also be considered as the  
coset
space $SU(k+1)/U(k)$. The defining representation of
$SU(k+1)$ is in terms of $(k+1)\times (k+1)$-matrices, which we may  
think of
as acting
on a $k+1$-dimensional vector space.
Let $t_A$ denote the generators of $SU(k+1)$
as matrices in this representation; we normalize them
by $\Tr (t_A t_B )=\half \delta_{AB}$.
The generators of $SU(k) \subset U(k)$ are then given by
$t_j $, $j =1, 2, ..., k^2-1$ and are matrices
which have zeros for the $k+1$-th row and column.
The generator corresponding to the $U(1)$ direction
of the subgroup $U(k)$ will be denoted by
$t_{k^2+2k}$. As a matrix,
\vspace{-3mm}
\beq
t_{k^2+2k} = {1\over \sqrt{2k(k+1)}}\left[
\begin{matrix}
1&0&\cdots&0&0\\
0&1&\cdots&0&0\\
\cdots&\cdots&\cdots&1&0\\
0&0&\cdots&0& - k
\end{matrix}
\right]\, .\label{1.7}
\eeq
We can use a general element of $SU(k+1)$, denoted by
$g$, to parametrize ${\bf CP}^k$, by making the identification
$g \sim gh$, where $h \in U(k)$.
In terms of $g$, the one-form $A$ is given by
\vspace{-3mm}
\beqar
A &=& i \sqrt{2k \over k+1}~ \Tr ( t_{k^2+2k} g^{-1} dg  
)\nonumber\\[1mm]
&=&-i g^*_{k+1,\alpha} dg_{\alpha , k+1}\ .
\label{1.8}
\eeqar
If we identify the normalized $u$ by
$u_\alpha = g_{\alpha , k+1}$, we see that this agrees with
(\ref{1.1}).
On the group element $g\in SU(k+1)$, considered as a $(k+1) \times
(k+1)$-matrix,
we can define left and right $SU(k+1)$ actions by
\beq
{\hat{L}}_A \,g = t_A \,g\,, \qquad\quad {\hat{R}}_A\,g = g\,t_A \ .
\label{1.9}
\eeq
If we denote the group parameters in $g$ by $\vf^i$, then we can write,  
in
general,
\beq
g^{-1} dg = -i t_A E^A_B \,d\vf^B, \qquad\quad  dg g^{-1} =
-it_A {\tilde E}^A_B\, d\vf^B\ .
\label{1.9a}
\eeq
For functions of $g$, the right and left translations are represented  
by the differential operators
\beq
{\hat R}_A = i (E^{-1})^B_A {\del \over \del \vf^B}\ , \qquad\quad
{\hat L}_A = i ({\tilde E}^{-1})^B_A {\del \over \del \vf^B}\ .
\label{1.9b}
\eeq

The action which we shall quantize is given by
\beq
{\mathcal S} = i n \sqrt{2k \over k+1} \int dt~\Tr\, ( t_{k^2+2k} g^{-1}
{\dot g} )\ .
\label{1.10}
\eeq
Since $\Tr (t_{k^2+2k} t_j) =0$ for generators $t_j$ of $SU(k)$, this
action is invariant
under $g \rightarrow gh$, for $h \in SU(k)$. For the $U(1)$  
transformations
of the
form $\exp ( i t_{k^2 +2k} \theta )$, the action changes by a boundary  
term;
the equations of motion are not affected and the classical theory is  
thus
defined on
$SU(k+1)/U(k)$, as needed.  In quantizing the theory, we observe that  
there is
no coordinate corresponding to the $SU(k)$ directions; the corresponding
canonical momenta
are constrained to be zero. Further, the canonical momentum  
corresponding to
the angle $\theta$ for the $U(1)$ direction is given by
$- nk /\sqrt{2k(k+1)}$. The states in the
quantum theory must thus obey the conditions
\vspace{-3mm}
\beqar
{\hat R}_j \,\Psi &=&0\,,\qquad\quad  j=1,\cdots , k^2 -1\,, \nonumber  
\\[1mm]
{\hat R}_{k^2 + 2k} \,\Psi&=& -nk\, {1 \over {\sqrt{2k(k+1)}}}\,\Psi\ .
\label{1.11}
\eeqar

Another way to see the last condition is to notice that, under
$g \rightarrow gh$, $h =\exp (i t_{k^2+2k} \theta )$,
the action changes by
\beq
\Delta {\mathcal S}= - {nk \over \sqrt{2k(k+1)}}\,\Delta  
\theta\label{1.12}
\eeq
leading to the requirement
\vspace{-3mm}
\beq
\Psi (gh ) = \Psi (g) \exp \left( -i\,{ nk \over \sqrt{2k(k+1)}}\,  
\theta \right)
\label{1.13}
\eeq
for wave functions $\Psi (g)$. (This also shows that the wave functions  
are not
genuine functions on ${\bf CP}^k$, but rather they are sections
of a $U(1)$ bundle on ${\bf CP}^k$.)

We will now consider these wave functions in some more detail.
A basis of functions on $SU(k+1)$
is given by the Wigner $\mathcal{D}$-functions which are the
matrices corresponding to the
group elements in a representation
$J$
\beq
\D^{(J)}_{L,R}(g) = \la J ,L_i \vert~ {\hat g}~\vert J, R_i \ra \label  
{1.14}
\eeq
where $L_i,~R_i$ stand for two sets of quantum numbers specifying the
states on which the
generators act, for left and right $SU(k+1)$ actions on $g$,  
respectively.
The
quantum numbers $R_i$ in (\ref{1.14}) must be constrained by
the conditions (\ref{1.11}). Thus the state $\vert J, R_i\ra$
corresponds to an $SU(k)$ singlet with a specific $U(1)$ charge
given by (\ref{1.11}). In addition to these conditions, we must recall
that $g$ parametrizes the whole phase space, and so the Wigner functions
depend on all phase space coordinates, not just half of them.
To eliminate half of them, we first define the derivatives on the phase  
space.

There are $2k$ right generators of $SU(k+1)$ which are not in
$U(k)$; these can be separated into $t_{+i}$ which are
of the raising type and $t_{-i}$ which are of the lowering
type. The derivatives on ${\bf CP}^k$
can be identified with these ${\hat R}_{\pm i}$ right rotations on $g$.
The ${\hat R}_{-i}$ commute among themselves, as
do the ${\hat R}_{+i}$'s; for the commutator between them we have
\beqar
[ {\hat R}_{+i}, {\hat R}_{-j}] &=&
i f_{ij}^a ~{\hat R}_a + \delta_{ij} \sqrt{ 2 (k+1) \over k}~~{\hat
R}_{k^2+2k} \nonumber\\
&=& -n ~\delta_{ij}
\label{1.15}
\eeqar
where ${\hat R}_a$ is a generator of
$SU(k)$ transformations, $f_{ij}^a$ are the appropriate
structure constants, and in the second line, we give the values
when acting on wave functions obeying (\ref{1.11}).
The derivatives thus split into conjugate pairs, analogously to
the creation and annihilation operators. Because of this,
the requirement that the wave functions
should not depend on half of the phase space coordinates can be taken as
\beq
{\hat R}_{-i} \Psi =0\ .\label{1.16}
\eeq
(In the geometric quantization approach to the action (\ref{1.10}) and  
the
construction of the Hilbert space, this is the so-called polarization
condition; for general works
on geometric quantization, see \cite{geomquant}.)
Based on the requirements (\ref{1.11}) and (\ref{1.16}),
we see that a basis for the wave functions is given by
the Wigner functions
corresponding to  irreducible
$SU(k+1)$
representations $J$, where right state $\vert J, R_i\ra$ is an $SU(k)$  
singlet,
has a value of $ -nk /\sqrt{2k(k+1)}$ for ${\hat R}_{k^2 +2k}$,
and further it must be a highest weight state, so that
(\ref{1.16}) holds.

Representations which contain an $SU(k)$ singlet, with the appropriate  
value
of ${\hat R}_{k^2+2k}$,
can be labeled by two
integers
$J = (p, q)$ such that $p-q = n$. The highest weight condition requires
  $p =n,~~q=0$. These are completely symmetric representations.
The dimension of this representation $J = (n, 0)$ is
\vspace{-5mm}
\beq
dim J = {{(n+k)!} \over {n! k!}} \equiv N\ ,
\label{1.17}
\eeq
$N$ expresses the number of states in the Hilbert space upon  
quantization.
Notice that $n$ has to be an integer for this procedure to go through;  
this
requirement
of integrality is the Dirac-type quantization condition mentioned  
earlier.

A basis for the
wavefunctions on ${\bf CP}^k$ can thus be written as
\beq
\Psi^{(n)}_m (g) = \sqrt{N}~\D^{(n)}_{m, -n} (g)\ . \label{1.18}
\eeq
We denote the fixed state for the right action on the Wigner
function above as $-n$, indicating that the eigenvalue for
${\hat R}_{k^2 + 2k}$ is $-nk / \sqrt{2k(k+1)}$ as in (\ref{1.11}).
The index $m$
specifies the state in this basis for the Hilbert space.
The Wigner $\D$-functions obey the orthogonality condition
\beq
\int d\mu (g) ~\D^{*(J)}_{m,k} (g)~\D^{(J')}_{m',k'}
(g) ~=~ \delta^{J J'}~{\delta_{mm'}\delta_{kk'}\over dim J}
\label{1.18a}
\eeq
where $d\mu (g)$ is the Haar measure on the group
$SU(k+1)$; we normalize it so that $\int d\mu (g) =1$.
Specializing to our case, we see that the wave functions (\ref{1.18})
are normalized since
\beq
\int d\mu (g) ~\D^{*(n)}_{m,-n} (g)~\D^{(n)}_{m',-n}
(g) ~=~ {\delta_{mm'}\over N}\ .
\label{1.19}
\eeq
Strictly speaking, the integration for the wave functions
should be over the manifold ${\bf CP}^k = SU(k+1)/U(k)$.
The measure should be the Haar measure for
$SU(k+1)/U(k)$; however,
we can integrate over the whole group since the integrand is
$U(k)$ invariant.

We now use the notation $ u_{\alpha}\equiv g_{\alpha, k+1}$;
In terms of the
$u_{\alpha}$'s, the Wigner $\D$-functions are of
the form
$\D \sim u_{\alpha_1} u_{\alpha_2}\cdots u_{\alpha_n}$.
Using the local complex coordinates introduced in (\ref{1.4}),
the wave functions (\ref{1.18}) are
\beqar
\Psi^{(n)}_m &=& \sqrt{N}~ \D^{(n)}_{m, -n}(g)\ , \nonumber\\
\D^{(n)}_{m, -n}(g) &=& \left[ {n! \over i_1! i_2! ...i_k!
(n-s)!}\right]^\half ~ {z_1^{i_1} z_2^{i_2}\cdots z_k^{i_k}\over
(1+\bz \cdot z )^{n \over 2}}\ , \label{1.20}\\[2mm]
s &=& i_1 +i_2 + \cdots +i_k \ . \nonumber
\eeqar
Here $0\leq i_l \leq n$, $0\leq s \leq n$.
The condition ${\hat R}_{-i} \D^{(n)}_{m,-n}=0$ is a holomorphicity
condition and this is reflected in the fact that the
wave functions are holomorphic
in $z$'s, apart from certain overall factors.
The states (\ref{1.20}) are coherent states for ${\bf CP}^k$  
\cite{perel}.
The inner product for the $\Psi$'s may be written in these coordinates
as
\beqar
\la \Psi \vert \Psi' \ra &=& \int d\mu ~\Psi^* ~\Psi'\ ,\nonumber\\[1mm]
d\mu &=& {k!\over \pi^k}{d^k z d^k\bz \over (1+\bz \cdot z )^{k+1}}\ .
\label{18}
\eeqar

At this point, we are able to define more precisely what we mean by
fuzzy ${\bf CP}^k$ \cite{NR1, grosse, bal1}.
Functions on fuzzy ${\bf CP}^k$ will correspond to
matrices acting on the $N$-dimensional Hilbert space given by the basis
(\ref{1.18}). They are thus $N\times N$-matrices and the composition law
is matrix multiplication. The composition law is not commutative and
so this corresponds to a noncommutative version of ${\bf CP}^k$.
There are $N^2$ independent elements for an arbitrary $N\times  
N$-matrix;
thus there are $N^2$ independent ``functions'' possible on fuzzy
${\bf CP}^k$ at finite $N$ (or finite $n$). What we need to show is  
that these
functions are in one-to-one correspondence with functions on the
usual commutative ${\bf CP}^k$, as $n \rightarrow \infty$. Further, in
this limit, the matrix product of two matrices tend to the
ordinary commutative product of the corresponding functions.
The first step is to define the symbol corresponding to any matrix;
the symbol is an ordinary function on commutative ${\bf CP}^k$
to which the matrix approximates in the large $n$ limit.
The matrix product can then be represented in terms of symbols
by a deformation of the ordinary
product, known as the star product. We now turn to a discussion of these
concepts and their properties.

\subsection{Star products, commutators and Poisson brackets}

Let $\hat{A}$ be a general matrix acting on the $N$-dimensional
Hilbert space generated by the basis (\ref{1.18})
with matrix elements $A_{ms}$.
We define the symbol corresponding to $A$ as the function
\beqar
A(g) &=&A(z, \bz )= \sum_{ms}\D^{(n)}_{m, -n}
(g) A_{ms} \D^{*(n)}_{s, -n}(g)\nonumber\\
  &=& \la -n\vert {\hat g}^T {\hat A} {\hat g}^* \vert -n\ra
\label{21}
\eeqar
where $\vert -n\ra = \vert J=n, -n\ra$.
We are interested in the
symbol corresponding to the product of two matrices
$A$ and $B$.
This can be written as
\beqar
(AB)(g)&=& \sum_r A_{mr} B_{rs} \D^{(n)}_{m, -n}
(g) \D^{*(n)}_{s, -n}(g) \nonumber\\
&=& \sum_{rr'p} \D^{(n)}_{m, -n}(g) A_{mr}
~\D^{*(n)}_{r,p}(g) \D^{(n)}_{r',p}(g)~
B_{r's} \D^{*(n)}_{s, -n}(g)
\label{22}
\eeqar
using $\delta_{rr'}=\sum_p
\D^{*(n)}_{r,p}(g)
\D^{(n)}_{r',p}(g)$.
The term with $p =- n$ on the right hand side
of (\ref{22}) gives the product of the symbols for $A$ and $B$.
The terms with $p > -n$ may be written using raising
operators as
\beq
\D^{(n)}_{r,p}(g) = \left[{(n-s)!\over n! i_1! i_2! \cdots i_k!}
\right]^\half {\hat R}_{+1}^{i_1}~{\hat R}_{+2}^{i_2} \cdots
{\hat R}_{+k}^{i_k} ~\D^{(n)}_{r,-n}(g)\ .
\label{23}
\eeq
Here $s=i_1+i_2+\cdots +i_k$
and the $t_{k^2+2k}$-eigenvalue for the state
$\vert p\ra$ is $(-nk +sk +s )/\sqrt{2k(k+1)}$.
Since ${\hat R}_{+i} \D^{*(n)}_{s, -n} =0$,
we can also write
\beq
\biggl[{\hat R}_{+i}\D^{(n)}_{r',-n}(g)\biggr] B_{r's} \D^{*(n)}_{s,  
-n} (g)
= \biggl[{\hat R}_{+i}
\D^{(n)}_{r',-n}
B_{r's} \D^{*(n)}_{s, -n} (g)\biggr] = {\hat R}_{+i} B(g)\ .
\label{24}
\eeq
Further keeping in mind that ${\hat R}_+^* = - {\hat R}_-$,
we can combine (\ref{22}-\ref{24}) to get
\beqar
(AB)(g) &=& \sum_s (-1)^s \left[ {(n-s)! \over n! s!}\right]
\sum_{i_1+i_2+\cdots +i_k=s}^n~
{s! \over i_1! i_2! \cdots i_k!}\nonumber\\[1mm]
&&\times{\hat R}_{-1}^{i_1} {\hat R}_{-2}^{i_2}
\cdots {\hat R}_{-k}^{i_k} A(g)
{\hat R}_{+1}^{i_1} {\hat R}_{+2}^{i_2}\cdots {\hat R}_{+k}^{i_k}
  B(g)\nonumber\\
&\equiv& A(g) * B(g)
\label{25}
\eeqar
The right hand side of this equation is what is known as the
star product for functions on ${\bf CP}^k$. It has been written down
in different forms in the context of noncommutative
${\bf CP}^k$ and related spaces \cite{bal1, alex}; our argument here
follows the presentation in \cite{nair}, which gives a simple and  
general
way of
constructing star products.
The first term of the sum on the right hand side gives $A(g) B(g)$,
successive terms involve derivatives and are down by powers of
$n$, as $n\rightarrow \infty$. For the symbol corresponding
to the commutator of $A, ~B$, we have
\beq
\bigl( [A,B]\bigr)(g) = -{1\over n}
\sum_{i=1}^{k}({\hat R}_{-i} A ~{\hat R}_{+i} B - {\hat R}_{-i} B~ {\hat
R}_{+i} A )
~+~ {\mathcal O} (1/n^2)\label{26}
\eeq

The K\"ahler two-form on ${\bf CP}^k$
may be written as
\beqar
\Omega &=&-i \sqrt{2k \over k+1}
\Tr \bigl( t_{k^2+2k} ~g^{-1}dg\wedge g^{-1}dg \bigr)\nonumber\\
&=& -{1\over 4} \sum_{i=1}^k~\left( E^{+i}_C E^{-i}_D - E^{-i}_C  
E^{+i}_D \right)
~d\vf^C \wedge d\vf^D
\nonumber\\
&=& -{1\over 4} \sum_{i=1}^k~\ep_{M_i N_i} E^{M_i}_C E^{N_i}_D
~d\vf^C \wedge d\vf^D
\nonumber\\
&\equiv& {1\over 2} \Omega_{CD} ~ d\vf^C \wedge d\vf^D
\label{27}
\eeqar
$\ep_{M_i N_i}$ is equal to $1$ for $m_i =+i, N_i =-i$ and is equal to  
$-1$
for $M_i =-i , N_i =+i$.

Functions $A,~B$ on ${\bf CP}^k$ obey the condition
${\hat R}_\alpha A = {\hat R}_\alpha B =0$, where ${\hat R}_\alpha$
(with the index $\alpha$)
is any generator of the subgroup $U(k)$.
With this condition, we find
\beq
i \sum_i \ep^{M_i N_i} (\hat{R}_{M_i} A) (E^{-1})^C_{N_i} ~\Omega_{CD}
= - {\del A \over \del \vf^D}
\label{28}
\eeq
The Poisson bracket of $A, ~B$, as defined
by $\Omega$, is
thus given by
\beqar
\{ A, B\} &\equiv& (\Omega^{-1})^{MN}
{\del A \over {\del \vf^M}} {\del B \over {\del \vf^N}}
\nonumber\\
&=& i \sum_{i=1}^{k}\left( \hat{R}_{-i} A~ \hat{R}_{+i} B ~-~
\hat{R}_{-i}B ~\hat{R}_{+i}A\right)
\label{29}
\eeqar
Combining with (\ref{26}), we find
\beq
\left( [A,B]\right) (g) = {i\over n} \{ A, B\} ~+~ {\mathcal O}(1/n^2)
\label{30}
\eeq
This is the general correspondence of commutators and Poisson brackets,
here realized for the specific case of
${\bf CP}^k$.
If desired, one can also write the Poisson bracket in terms of the local
coordinates $z ,\bz$ introduced in (\ref{1.4}).
The relevant expressions are
\beqar
\{ A, B\}&=& i (1+\bz \cdot z )
\Biggl( {\del A \over \del z^i}
{\del B\over \del \bz^i} - {\del A \over \del \bz^i }{\del B\over\del  
z^i}
+z\cdot {\del A\over \del z}
\bz \cdot {\del B\over\del\bz} -\bz\cdot {\del A \over
   \del\bz}
z\cdot {\del B\over \del z} \Biggr)\,.\nonumber\\
\label{31}
\eeqar

\vspace{-4mm}

The trace of an operator ${\hat A}$ may be written as
\beqar
\Tr {\hat A} &=& \sum_m {A}_{mm} = N~\int d\mu (g)
\D^{(n)}_{m, -n} ~A_{mm'}~
\D^{*(n)}_{m', -n}\nonumber\\
&=& N \int d\mu (g)~ A(g)\ .
\label{32}
\eeqar
The trace of the product
of two operators $A, B$ is then given by
\beq
\Tr {\hat A} {\hat B} ~=~ N \int d\mu (g) ~A(g)*B(g)\ .
\label{33}
\eeq

\subsection{The large $n$-limit of matrices}

We now consider the symbol for the product ${\hat T}_B {\hat A} $ where
${\hat T}_B$ are the generators of $SU(k+1)$,
viewed as linear operators on the states
in the representation $J$. Using the formula (\ref{21}), it can be
simplified as
follows.
\beqar
({\hat T}_B {\hat A} )_{\alpha\beta} &=& \la \alpha \vert~ {\hat g}^T~
{\hat T}_B ~{\hat A}~ {\hat g}^*
~\vert \beta \ra \nonumber\\[1mm]
&=& S_{BC} ~\la \alpha \vert ~{\hat T}_C ~{\hat g}^T ~{\hat A}~ {\hat  
g}^*
~\vert \beta \ra
\nonumber\\[1mm]
&=& S_{Ba} (T_a)_{\alpha\gamma}~ \la \gamma \vert ~{\hat g}^T ~{\hat A}~
{\hat g}^* ~\vert\beta\ra
+ S_{B+i} ~\la \alpha \vert~ {\hat T}_{-i}~ {\hat g}^T ~{\hat A}~
{\hat g}^* ~\vert \beta \ra\nonumber\\[1mm]
&&\hskip .05in
+ S_{B~k^2+2k}~ \la \alpha \vert~ {\hat T}_{k^2+2k}~ {\hat g}^T~ {\hat
A} ~{\hat g}^* ~\vert \beta \ra
\nonumber\\[1mm]
&=& {\mathcal L}_{B\alpha \gamma}~ \la \gamma \vert ~{\hat g}^T ~{\hat  
A}~
{\hat g}^* ~\vert \beta\ra
\nonumber\\[1mm]
&=& {\mathcal L}_{B\alpha \gamma}~ A(g)_{\gamma\beta}
\label{42}
\eeqar
where we have used ${\hat g}^T {\hat T}_B {\hat g}^* = S_{BC} {\hat  
T}_C$,
$S_{BC} = 2 \Tr (g^T t_B g^* t_C)$.
(Here $t_B, t_C$ and the trace are in the fundamental representation of
$SU(k+1)$.)
${\mathcal L}_B $ is defined as
\beq
{\mathcal L}_{B\alpha\gamma}=
-\delta_{\alpha\gamma}{nk\over \sqrt{2k(k+1)}}S_{B ~k^2 +2k}~
~ +~\delta_{\alpha\gamma} S_{B+i} {\hat{{\tilde R}}}_{-i}
\label{43}
\eeq
and ${\hat{{\tilde R}}}_{-i} $ is a differential operator
defined by $\hat{{\tilde R}}_{-i} g^T
= T_{-i} g^T $. (This can be related to ${\hat R}_{-i}$ but it is
immaterial here.)
We have also used the fact that the states $\vert \alpha\ra$, $\vert  
\beta\ra$
are $SU(k)$-invariant.
By choosing ${\hat A}$ as a product of $\hat{T}$'s, we can extend
the calculation of the symbol for any product of $\hat{T}$'s
using equation (\ref{42}). We find
\beq
({\hat T}_A {\hat T}_B \cdots {\hat T}_M)_{\alpha \beta}
= {\mathcal L}_{A\alpha\gamma_1} {\mathcal L}_{B\gamma_1 \gamma_2  
}\cdots
{\mathcal L}_{M \gamma_r \beta} \cdot 1\ .
\label{44}
\eeq

A function on fuzzy ${\bf CP}^k$ is an $N\times N$-matrix.
It can be written as a linear combination of products of ${\hat T}$'s
and by using the above formula, we can obtain its large $n$ limit.
When $n$ becomes very large, the term that dominates in ${\mathcal L}_A$
is $S_{A~k^2+2k}$.
We then see that for any matrix function we have the relation,
$F({\hat T}_A ) \approx F(S_{A~k^2+2k})$.

We will now define a set of ``coordinates'' $X_A$ which are $N\times
N$-matrices
by
\beq
X_A = -{1\over \sqrt{C_2(k+1,n)}}~{\hat T}_A\label{45}
\eeq
where
\beq
C_2( k+1 ,n)  = {n^2k^2\over 2k (k+1)} + {nk\over 2}
\label{45a}
\eeq
is the quadratic Casimir value for the
symmetric rank $n$ representation.
$X_A$ will be considered as coordinates of fuzzy ${\bf CP}^k$ embedded
in ${\bf R}^{k^2+2k}$. In the large $n$ limit, we evidently have
$X_A \approx S_{A ~k^2+2k} = 2 \Tr (g^T t_A g^* t_{k^2+2k})$.
By its definition, $S_{A ~k^2+2k}$ obey algebraic constraints
which can be verified to be the correct ones for
describing ${\bf CP}^k$ as embedded in
${\bf R}^{k^2+2k}$.

\subsection{The symbol and diagonal coherent state representation}

The states we have constructed are the coherent states for the
${\bf CP}^k$ and we have an associated diagonal coherent state
representation \cite{perel},\cite{Randjbar-Daemi:1992zj}. Notice that  
the
states have the expected holomorphicity
property that coherent states have. In fact, the condition (\ref{1.16})
is the statement of holomorphicity; explicitly, the wave functions
in (\ref{1.20}), apart from the
prefactor involving
$(1 +\bz \cdot z )^{-{n\over 2}}$, involve only $z_i$ and not $\bz_i$.
We now show that the
symbol of an operator
is related to, but is not exactly, the expectation value of the operator
in this coherent state representation.
The first step towards this is
the Wigner-Eckart theorem, which is a standard result, and can also be  
seen
easily as follows.

Let $F_\alpha$ be a tensor operator belonging to the representation $r$.
We then have
\beq
{\hat g} F_\alpha {\hat g}^\dagger = \D^{(r)}_{\beta ,\alpha} F_\beta\ .
\label{46}
\eeq
We can now write
\beqar
(F_\alpha )_{km} &=& \la n, k\vert F_\alpha \vert n,  
m\ra\nonumber\\[1mm]
&=& \la n, k\vert {\hat g}^\dagger {\hat g} F_\alpha
{\hat g}^\dagger {\hat g} \vert n, m\ra\nonumber\\[1mm]
&=& \sum_{p,q,\beta}  \D^{(n)}_{k,p}(g^\dagger )~ \D^{(n)}_{q,m}(g)
\D^{(r)}_{\beta,\alpha}(g)~\la n, p\vert F_\beta \vert n, q\ra\ .
\label{46a}
\eeqar
In this expression, we can combine the product of
the representations by the Clebsch-Gordon theorem;
the Clebsch-Gordon coefficients $\la j, p\vert r,\beta ; n,q\ra$
for reduction of product
representations are defined by
\beq
\vert r,\beta;  n,q\ra = \sum_{j,p} \la j,p\vert r,\beta ; n,q\ra  
~\vert j,
p\ra\label{46b}
\eeq
where the sum is over all representations
which can be obtained by the  product of representations $r$ and $n$,
and over the states within each such representation.
Using this result and integrating both sides of equation (\ref{46a})
over all $g$, we get
the Wigner-Eckart theorem
\beq
(F_\alpha )_{km} = \la n,k \vert r,\alpha ; n, m\ra  ~\la\!\la F  
\ra\!\ra
\label{47}
\eeq
where the reduced matrix element $\la\!\la F \ra\!\ra$ is given by
\beq
\la\!\la F \ra\!\ra = \sum_{p,q,\beta} {1\over N} \, \la r ,\beta ; n,  
q\vert
n , p\ra ~\la n, p\vert F_\beta \vert n, q\ra
\label{48}
\eeq
where $N$ is the dimension of the representation labeled by $n\,$.

Now, from the definition of the Clebsch-Gordon coefficient in  
(\ref{46b}),
we have the relation
\beq
\int d\mu (g) \D^{(n)}_{k,p}(g^\dagger )~ \D^{(n)}_{q,m}(g)
\D^{(r)}_{\beta,\alpha}(g)
={1\over N}\, \la r,\beta ; n, q\vert n, p\ra ~ \la n, k \vert r,\alpha  
; n, m\ra \ .
  \label{49}
\eeq
We can put $\beta =0$, $p =q =-n$ in this equation and use (\ref{47})  
to obtain
\beqar
&&\la\!\la F \ra\!\ra ~\la n, k \vert r,\alpha ; n,m\ra  ~ \la r ,0; n,  
-n\vert
n, -n\ra \nonumber\\
&&\hskip .8in = N \int d\mu (g) ~ \D^{(n)*}_{-n,k}(g^\dagger )~
\D^{(n)}_{-n,m}(g)
\la\!\la F \ra\!\ra  \D^{(r)}_{0,\alpha}(g)\nonumber\\
&&\hskip .8in = N\int d\mu (g) \D^{(n)*}_{k, -n} (g^T) \D^{(n)}_{m,  
-n}(g^T)
\la\!\la F \ra\!\ra \D^{(r)}_{\alpha, 0}(g^T)\nonumber\\
&&\hskip .8in = \int d\mu (g)~\Psi^{(n)*}_k \left[  \la\!\la F \ra\!\ra
\D^{(r)}_{\alpha, 0}(g)\right]  \Psi^{(n)}_m
\label{49a}
\eeqar
where, in the last step, we made a change of variable $g \rightarrow  
g^T$,
and used the definition of states (\ref{1.20}).
We now define a function $f_\alpha (g)$ by
\beq
f_\alpha (g) ~\la r,0; n,-n\vert n,-n\ra = \D^{(r)}_{\alpha, 0} (g)
\la\!\la F\ra\!\ra\ .
\label{50}
\eeq
We can then rewrite (\ref{49a}) as
\beqar
\int d\mu (g) \Psi^{(n)*}_k ~f_\alpha (g) ~\Psi^{(n)}_m
&=&\la n,k\vert r,\alpha ; n,m \ra~ \la\!\la F \ra\!\ra \nonumber\\
&=& (F_\alpha )_{km}\ .\label{51}
\eeqar
We see that the matrix element of $F_\alpha$ can be reproduced by
the expectation value of a function $f_\alpha (g)$. This is the diagonal
coherent state
representation.
The result is easily extended to any matrix since we can always write  
it as
a linear combination of tensor operators which have definite  
transformation
properties.
Thus
\vspace{-3mm}
\beq
\la k\vert {\hat F} \vert m\ra = \int d\mu (g) ~\Psi^{(n)*}_k~f(g)
~\Psi^{(n)}_m
\label{52}
\eeq
or, equivalently,
\vspace{-3mm}
\beq
{\hat F} = \int d\mu (g) ~ \vert g \ra~ f(g)~ \la g \vert
\label{53}
\eeq
where the states $\vert g\ra$ are defined by
$\la k \vert g\ra = \Psi^{(n)*}_k$, $ \la g\vert m\ra = \Psi^{(n)}_m$.

The function $f(g)$ is not the same as the symbol for $F$. In fact, the
symbol may be written as
\beqar
F(g) &=& \sum_{km} \D^{(n)}_{k, -n}(g)~ F_{km} ~\D^{(n)*}_{m,  
-n}(g)\nonumber\\
&=& N \sum_{km} \int_h \D^{(n)}_{k, -n}(g) \D^{(n)*}_{k, -n}(h) ~ f(h) ~
\D^{(n)}_{m, -n} (h)~ \D^{(n)*}_{m, -n}(g)\label{54}\\
&=& \int_h \D^{(n)*}_{-n,-n} (g^\dagger h) ~ f(h)~ \D^{(n)}_{-n,
-n}(g^\dagger h)\nonumber\\
&=& \int_u \D^{(n)*}_{-n,-n} (u) ~ f(gu)~ \D^{(n)}_{-n, -n}(u)\  
.\nonumber
\eeqar
Since $F$ can be written as a combination of tensor operators, we have
$f(g) = \sum_{r, \alpha} C^r_\alpha \D^{(r)}_{\alpha, 0}$, for some  
coefficient
numbers $C^r_\alpha$. Using this in the above equation,
we get
\vspace{-2mm}
\beqar
F(g) &=& \sum_{r, \alpha} C^r_\alpha \D^{(r)}_{\alpha ,0} (g) ~\vert
{\mathcal C}\vert^2\ ,
\label{55}\\
{\mathcal C}&=&\la n, -n\vert r,0; n, -n\ra \ .\nonumber
\eeqar
In the large $n$ limit, ${\mathcal C}\rightarrow 1$, and
the symbol and the function $f$ coincide.

\setcounter{equation}{0}
\section{Special cases}

It is instructive at this point to consider some special cases.
\vspace{-3mm}
\subsection{The fuzzy two-sphere}

This is one of the best-studied cases
\cite{madore2}. Since $S^2 \sim {\bf CP}^1 = SU(2)/U(1)$, this is the  
special
case of $k=1$ in our analysis. In this case, the representations of  
$SU(2)$ are
given by the standard angular momentum theory.
Representations are labeled by the maximal angular momentum $j =  
{n\over 2}$,
with $N =2j+1 =n+1$.
The generators of the group are the angular momentum matrices, and one  
may
identify the
coordinates of fuzzy $S^2$ by $X_i = J_i /\sqrt{j(j+1)}$.
At finite $n$, the coordinates do not commute,
\beq
[X_i , X_j ] = {i\over \sqrt{j(j+1)}}\, \epsilon_{ijk} X_k\ . \label{56}
\eeq
We can parametrize an element of $SU(2)$ as
\beq
g = {1\over \sqrt{(1+\bz z)}} \left[ \begin{matrix}\bz &~1\cr
-1&~z\cr\end{matrix}\right]\,.
\label{57}
\eeq
For $S_{i3}(g)$ we then find
\beqar
S_{13}= - {z +\bz \over (1+z\bz )}\ , &\hskip .3in& S_{23} = -i \,{{z  
-\bz}
\over (1+z\bz)}\ ,
\hskip .3in
S_{33} = {z\bz -1 \over z\bz +1}\ .
\label{58}
\eeqar
At the matrix level, we have $X_i X_i =1$; in the large $n$ limit,
$X_i \approx S_{i3}$, which also obey the same condition.
$z, \bz$ are the local complex coordinates for the sphere.

A function on fuzzy $S^2$ is an $N\times N$-matrix, so,
at the matrix level, there are $N^2 = (n+1)^2$ independent  
``functions''.
On the smooth $S^2$, a basis for functions is given by the spherical  
harmonics,
labeled by the integer $l = 0, 1, 2,...$.
There are $(2l+1)$ such functions for each value of $l$.
If we consider a truncated set of functions with a maximal value
of $l$ equal to $n$, the number of functions is
$\sum_0^n (2l+1) = (n+1)^2$. Notice that this number coincides with
the number of ``functions'' at the  matrix level.
By using the relation $X_i \approx S_{i3}$, we can see that the  
functions
involved correspond to products of $S_{i3}$ with up to
$n$ factors. These are in one-to-one correspondence with the spherical
harmonics, for $l=0, 1, 2$, etc., up to $l=n$, since $S_{i3}$ has  
angular
momentum $1$.
Thus we see that the set of functions at the matrix level will go over  
to the
set of functions on the smooth $S^2$ as $n \rightarrow \infty$.

Fuzzy $S^2$ may thus be viewed as a regularized version of the smooth
$S^2$ where we impose a cut-off on the number of modes of a function.
$n$ is the regulator or cut-off parameter.

\subsection{Fuzzy ${\bf CP}^2$}

This corresponds to the case $k =2$ \cite{NR1, grosse, bal1}. The large  
$n$
limit of the
coordinates $X_A$ are $S_{A8} = 2 \Tr (g^T t_A g^* t_8)$.
It is easily checked that, in this limit, they obey the conditions
\beqar
X_A X_A &=& 1\ ,\nonumber\\
d_{ABC} X_B X_C &=& -{1\over \sqrt{3}}\, X_C\label{59}
\eeqar
where $d_{ABC} = 2\Tr t_A (t_B t_C + t_C t_B)$.
These conditions are well known to be the equations representing ${\bf  
CP}^2$
as embedded in ${\bf R}^8$.
Thus, in the large $n$ limit, our definition of fuzzy ${\bf CP}^2$ does
recover the
smooth ${\bf CP}^2$. One can impose these conditions at the level of  
matrices
to get a purely matrix-level definition of fuzzy ${\bf CP}^2$.

The dimension $N$ of the Hilbert space is in this case given by
$\half (n+1) (n+2)$. We may also consider the matrix functions which are
$N\times N$-matrices; they can be thought of as products of the ${\hat
T}$'s with up to $N-1$ factors.
There are $N^2$ independent functions possible. On the smooth ${\bf  
CP}^2$,
a basis of functions is given by (\ref{1.5a}). There are $d(2,l)
= (l+1)^3$ such functions for each value of $l$.
If we consider a truncated set, with values of $l$ going up to
$n$, the number of independent functions will be
\beq
\sum_0^n (l+1)^3 = {1\over 4} (n+1)^2 (n+2)^2 = N^2\, .
\label{60}
\eeq
It is thus possible to consider the fuzzy ${\bf CP}^2$ as a  
regularization
of the smooth ${\bf CP}^2$ with a cut-off on the number of modes of a  
function.
Since any matrix function can be written as a sum of products of ${\hat  
T}$'s,
the corresponding  large $n$ limit has a sum of
products of $S_{A8}$'s.
The independent basis functions are thus given by representations of
$SU(3)$ obtained from reducing symmetric products of the adjoint  
representation
with itself; they are the $\phi_l$'s of equation (\ref{1.5a}).
In fact, since the states are of the form $u_{\alpha_1} u_{\alpha_2}...
u_{\alpha_n}$, a general linear transformation is of the form
$M^{\alpha_1 ...\alpha_n}_{\beta_1 ...\beta_n}$. The traceless part of  
this
forms the irreducible representation $\phi_n$, the traces give lower
rank irreducible representations $\phi_l$, $ l<n$. We see that there is
complete agreement with the set of functions on ${\bf CP}^2$ with a  
cut-off
on the modes at $l =n$.

Since we can regard fuzzy ${\bf CP}^k$ as a regularization
of the smooth ${\bf CP}^k$ with a cut-off on the number of modes of a  
function,
one can use these fuzzy spaces to construct regulated field theories,
in much the same way that lattice regularization of field theories is  
carried out.
There are novel features associated with such a regularization; for  
example, the
famous (or notorious)
fermion doubling problem on the lattice can be evaded in an interesting  
way.
For these and other details, see \cite{fuzzfield}.

\setcounter{equation}{0}
\section{Fields on fuzzy spaces}

In this section we will briefly consider how one can define a field  
theory on
a fuzzy space.

A scalar field $\Phi $ on a fuzzy space is obviously an $N\times  
N$-matrix
which can take arbitrary values. We may write $\Phi (X)$, indicating  
that it is
a function of the coordinate matrices $X_A$. For constructing an  
action, we
need
derivatives. From the general property (\ref{30}), we see that we can  
write
\beqar
[T_A , \Phi ] &\approx& - {i \over n}\, {nk \over \sqrt{2k (k+1)}}\, \{
S_{A~k^2+2k},
\Phi \}\nonumber\\[1mm]
&\equiv&-i D_A \Phi \ .\label{61}
\eeqar
On the left hand side of this equation
we have the matrix quantities while on the right hand side we have the
corresponding symbols.
$D_A$, as defined by this equation, are given by
\beqar
D_A &=& \sqrt{k \over 2(k+1)} ~ (1+\bz z) \Biggl[ \left({\del \over  
\del z^i}
+ \bz_i z\cdot {\del \over \del z}\right)S_{A~k^2+2k}  {\del \over \del
\bz^i}\nonumber\\
&&\hskip 1.45in- \left({\del \over \del \bz^i}
+ z_i \bz\cdot{ \del \over \del \bz}\right)S_{A~k^2+2k} t {\del \over  
\del z^i}
\Biggr]\,,
\label{62}
\eeqar
$D_A$ are derivative operators appropriate to the space we are  
considering.
For example, for the fuzzy $S^2$, we find
\beqar
D_1 &=& {1\over 2}\, ( \bz^2 \del_{\bz} +\del_z - z^2 \del_z
-\del_{\bz})\ ,\nonumber\\
D_2 &=& -{i\over 2}\, ( \bz^2 \del_{\bz} + \del_z +z^2 \del_z
+\del_{\bz})\ ,\label{63}\\[1mm]
D_3 &=& \bz \del_{\bz} - z \del_z \ . \nonumber
\eeqar
These obey the $SU(2)$ algebra, $[D_A ,D_B] = i \epsilon_{ABC} D_C$;
they generate the translations on the two-sphere. They are, in fact,  
the three
isometry transformations. This shows that we can define the derivative
of $\Phi$, at the matrix level, as the commutator $i [T_A, \Phi ]$,
which is the adjoint action of $T_A$ on $\Phi$.
The Laplace operator is then given by
$- \Delta \cdot \Phi = [T_A, [T_A, \Phi ]]$.

An example of the Euclidean action for a scalar field is then
\beq
{\mathcal S} = {1\over N} \Tr \biggl[  \Phi^\dagger  [T_A, [T_A, \Phi  
]]  +
V (\Phi )\biggr]
\label{64}
\eeq
where $V(\Phi )$ is a potential term which does not involve derivatives.

Another interesting class of fields is given by
gauge fields. Since the derivatives are given by the adjoint action
of the $T_A$, we can introduce a gauge field ${\mathcal A}_A$ and the  
covariant
derivative
\beq
-i \D_A \Phi = [T_A , \Phi ] + {\mathcal A}_A \Phi
\label{65}
\eeq
where ${\mathcal A}_A$ is a set of hermitian matrices. In the absence  
of the
the gauge field, we have the commutation rules $[T_A, T_B] =i f_{ABC}  
T_C$,
so that the field strength tensor ${\mathcal F}_{AB}$ may be defined by
\beq
-i {\mathcal F}_{AB} =  [ T_A +{\mathcal A}_A , T_B +{\mathcal A}_B] -
if_{ABC} (T_C +{\mathcal A}_C)\ .
\label{66}
\eeq
One can now construct a Yang-Mills type action for a gauge theory
as
\beq
{\mathcal S} = {1\over N}\, \Tr \biggl[ {1\over 4}\,{\mathcal F}_{AB}  
{\mathcal
F}_{AB}
\biggr]\,.\label{67}
\eeq

Starting with actions of the type (\ref{64}) and (\ref{67}), it is  
possible
to develop
the functional integral for the quantum theory of these fields and do
perturbation theory in terms of Feynman diagrams, etc. We will not do  
this
analysis here for two reasons. The analysis of field theories on fuzzy  
spaces
where the nontrivial geometry plays an important role has not yet been
developed to a great extent. We give some of the references which can  
point
the reader
to ongoing work \cite{fuzzfield}. Properties of field theories on flat
noncommutative spaces
(which we have not discussed here) has been more extensively  
investigated;
for this there are good reviews available \cite{douglas, others2}.

\setcounter{equation}{0}
\section{Construction of spheres}

We have discussed in some detail the complex projective spaces.
They are the spaces which emerge most naturally in any matrix  
construction.
The reason is simple. Matrices are linear operators on a Hilbert space
and so they are related to the quantization of a classical phase space.
Therefore, spaces which admit a symplectic structure are natural  
candidates
for fuzzification. Spheres, except for $S^2$ and products thereof, do  
not
fall into this
category. The construction of the spheres is thus more involved. The  
general
strategy has been to identify them as subspaces of suitable fuzzy spaces
and to introduce conditions restricting the functions to be on the  
sphere.

There is good reason to seek fuzzification of spheres, apart from the
general mathematical
interest in constructing them.
As we mentioned before, one of the ways fuzzy spaces can be used
is that they provide a finite mode truncation of field theories.
Therefore, one may think of them as an alternative to the usual
lattice formulation of field theory which is necessary to formulate
field theories in a finite way and ask and answer questions about  
whether
they exist and so on. Further, they can be useful for numerical  
analyses of
field theories.
Four dimensions, of course, are the most interesting from this point of  
view;
however, ${\bf CP}^2$ is not the best, since the smooth ${\bf CP}^2$  
does
not have
a spin structure. (It can have a so-called ${\rm spin}^c$ structure,  
with
an additional
$U(1)$ field, which we can take to be the "monopole potential"  given in
(\ref{1.1}) \cite{hawk}.
For a recent analysis of the solution of Dirac equation in the  
background of this field,
see \cite{Dvali:2001qr}.)
Fuzzification of $S^4$ would be very useful for this.

The method for the construction of
spheres is exemplified and illustrated by the case of fuzzy $S^1$.
We start with fuzzy $S^2 \sim {\bf CP}^1$; the modes at finite $n$
are the fuzzy versions of the spherical harmonics $Y^l_m (\theta ,\vf )$
with $l =0, 1$, etc.,
up to $l=n$. Take the highest spherical harmonic; this has $m$ values
$-n, -n+1,$ etc., up to $n$. The $\vf$-dependence of this function
corresponds to
modes $e^{im\vf}$ on $S^1$ for the same range of values for $m$.
If we take the large $n$ limit, we see that
this single spherical harmonic can give all the modes required for  
$S^1$.
So, one strategy, advocated in \cite{ram}, is to introduce a projection
operator which,
acting on a matrix $F$,
which may be viewed as a function on the fuzzy space,
retains only the highest mode. If we split $F$ as $F= F_+ +F_-$, where
$F_+$ is the part corresponding to the harmonics on $S^1$ and
$F_-$ the remainder, the projection operator $P$ is defined by
$P (F) = F_+$.
This can be done for higher dimensional spheres as well. The difficulty  
with
this approach is that the product of two such projected matrices will  
generate
the other unwanted modes, so we have to define the operation of  
multiplication
by $F * G = P( FG)$; the algebra is done before the projector is  
applied.
This product is not associative in general, so that one cannot interpret
the matrix functions on the fuzzy sphere as a linear transformation on a
Hilbert space; in the large $n$ limit, associativity is
recovered. Nevertheless, the lack of associativity limits the utility of
this approach.

A related idea starts with the question: what do we want to use the
fuzzy sphere for? If it is for the purpose of constructing field  
theories on it
and studying their behavior as $n \rightarrow \infty$, then a different
strategy is possible \cite{dolan1}.
It would be more practical, even for numerical simulations of theories,
to include all modes,
for example, all the spherical harmonics for all $l\leq n$, so that we  
do have
the algebra of functions on the bigger space, the fuzzy two-sphere in  
this
example.
One can then choose the action, so that all the unwanted modes have a
large contribution to the  Euclidean action. Such modes are then  
suppressed
and one gets a softer way of approaching the modes on the sphere we  
want.
In the example of $S^1$, notice that the quantity $h [n(n+1) - T^2]$,  
where
$T_A$ are the angular momentum generators, is positive for all
$l < n$ and is zero for $l =n$. Thus adding a term with this eigenvalue
would prejudice it against all modes $l<n$, and by taking the parameter
$h$ to be large, we can get a good approximation to the
circle $S^1$.
For a scalar field, such an action is given by
\beq
{\mathcal S} = {1\over n+1}\,\Tr \left[ {1\over 2}\, \Phi^\dagger
[T_3, [T_3 , \Phi ]] + {h \over 2}\, \Phi^\dagger [ n (n+1) - T^2  
]\cdot \Phi
\right]
\label{s1}
\eeq
where $T^2\cdot \Phi = [T_A , [T_A , \Phi]]$.

These ideas can be extended to higher dimensional spheres.
We have to start with suitable co-adjoint orbits to construct the
Hilbert space via quantization as we have done. For
spheres,
some of the useful co-adjoint orbits are \cite{dolan1}:
\begin{enumerate}
\item $SO(3)/ SO(2) \sim S^2$.
\item $SO(4)/[SO(2) \times SO(2)]~\sim S^2 \times S^2$.
(This can be related to $S^3 / {\bf Z}_2$ as we show below.)\\[-2mm]
\item $SO(5)/ [SO(3) \times SO(2)] ~\sim {\bf CP}^3 /{\bf Z}_2$.
(This can be used for the fuzzy version of $S^4$
utilizing the fact that ${\bf CP}^3$ is an $S^2$ bundle over $S^4$.
It can also be used to
approximate $S^3$ by prejudicing the action against the
unwanted modes as in the case of $S^1$.)\\[-2mm]
\item $SO(N+2) /[SO(N) \times SO(2)]$. (This can be used for higher  
dimensional
spheres in a similar way.)
\end{enumerate}

A matrix version of the four-sphere, which is useful in
the context of solutions to M(atrix) theory,
is worthy of special mention \cite{castelino}. For the four sphere,
we expect to have five matrices $X_\mu $, $\mu = 1, ..., 5$,
such that $X_\mu X_\mu =1$. This can be achieved by
using the Euclidean Dirac $\gamma$-matrices. There is only
one irreducible representation for the $\gamma$ matrices, so
to get a sequence of larger and larger matrices, one can take
tensor products of these,
\beq
X_\mu = ( \gamma_\mu \otimes 1 \otimes 1 ...\otimes 1~+~
1 \otimes \gamma_\mu \otimes 1 ...\otimes 1 ~+~\cdots )_{sym}
\label{s2}
\eeq
where the subscript $sym$ indicates symmetrization.

We now consider the fuzzy version of
$S^3/{\bf Z}_2$ which is related to $S^2 \times S^2$ \cite{NR2}.
The end result is not quite a sphere, but it is still an interesting  
example,
since there is an algebra of functions which has closure under
multiplication.
In this case,
$S^2 \times S^2$ plays a role, vis-\`a-vis $S^3/{\bf Z}_2$,
analogous to what ${\bf CP}^3$
does for
$S^4$.

In the smooth limit, the space
$S^3/{\bf Z}_2$ can be embedded in $S^2\times S^2$. The
latter space
can be described by $n^2 =1$, $m^2=1$,
$n =( x_1, x_2, x_3)$, $m =( y_1, y_2, y_3)$.
The space $S^3/{\bf Z}_2$ is now obtained by imposing
the further condition $n\cdot m = x_1y_1 +x_2 y_2 +x_3y_3
=0$. It is clear that any solution to these
equations gives an $SO(3)$ matrix $R_{AB}=
(\epsilon_{ABC}m_B n_C,
m_A, n_A)$. Conversely, given any element $R_{AB}\in
SO(3)$,
we can identify $n_A = R_{A3}$, $m_A =R_{A2}$. There are
other ways
to identify $(n,m)$ but these are equivalent to
choosing different sets of values
for the $SO(3)$ parameters; this statement can
be easily checked using the Euler angle parametrization.
What we have described is essentially the angle-axis
parametrization of rotations \cite{tung}.
Since the
space $S^2\times S^2$ has K\"ahler structure,
it is the simplest enlargement of space we can use
to
define coherent states.

We now turn to
the  fuzzy version
of $S^3/{\bf Z}_2$.
Consider $SU(2)\times SU(2)$, with generators
$T_A , T'_A$ and take a particular representation
where $l=l'$, so that we can think of $T, ~T'$ as
$(N\times N)$-matrices, $N=2l+1$.
Since the quadratic Casimirs $T^2 =T'^2 = l(l+1)$,
this gives
the standard realization of fuzzy $S^2 \times S^2$
\cite{madore2}.
As $l$ becomes large, we can use our result (\ref{43})
\beq
T_A \approx l ~2 \,\Tr \,(g^\dagger t_A g t_3) ,
\qquad\quad
T'_A \approx l ~ 2 \,\Tr\, ( g'^\dagger t_A g' t_3)
\label{s3}
\eeq
where $g, g'$ are $(2\times 2)$-matrices
parametrizing the two $SU(2)$'s.
All functions of $T, T'$ are similarly
approximated.
To get to a smaller space, clearly we need to
put an additional restriction which we will take
as the following.
An operator is considered admissible or physical if
it commutes with $T\cdot T'$, or equivalently
commutes with $(T-T')^2$ or $(T+T')^2$, i.e,
\beq
[ {\mathcal O}, (T-T')^2 ]=0\ .
\label{s4}
\eeq
It is easily seen that
the product of any two operators which obey this
condition will also obey the same condition, so this leads
to a closed algebra.
A basis for the vector space on which $T, T'$ act
is given by $\vert l m l m'\ra$ in the standard angular
momentum notation. We rearrange these into multiplets
of $J_A = T_A +T'_A$. For all
state within each irreducible representation
of the $J$-subalgebra labeled by $j$, $(T-T')^2$
has the same eigenvalue $4 l(l+1) - j (j+1)$.
Operators which commute with it are thus block diagonal,
consisting of all unitary transformations on
each $(2j+1)$-dimensional subspace.
There are $(2j+1)^2$ independent transformations
for each $j$-value putting them in one-to-one
correspondence
with the basis functions $\D^{j}_{a,b}(U)$ for an
$S^3$ described by the $SU(2)$ element $U$.
By construction, we get only integral values of
$j$, even if $l$ can be half-odd-integral,
so we certainly cannot
get $S^3$ in the large $l$ limit, only
$S^3/{\bf Z}_2$.

We can go further and ask how the condition
(\ref{s4}) can be implemented in the large
$l$ limit. This can be done by fixing the value of
$T\cdot T'$ to be any constant. Using (\ref{s3})
we find that this leads to
\beq
T\cdot T' \sim 2\, \Tr\, ( g'^\dagger g ~t_3 g^\dagger
g' t_3 )
\sim {\rm constant}\ .
\label{s5}
\eeq
This means that
\beq
g'^\dagger g = M \exp(it_3 \theta )
\label{s6}
\eeq
where $M$ is a constant $SU(2)$ matrix.
$\theta$ can be absorbed into $g$.
Since $T\cdot T'\sim
2\, \Tr \,( M t_3 M^\dagger t_3)$, we can take
$M = \exp (it_2 \beta_0)$ using the Euler
angle parametrization. We then find
\beqar
T_A &\sim& 2\, \Tr\, (g^\dagger t_A g t_3 )\ ,\nonumber\\[1mm]
T'_A &\sim& \cos\beta_0 ~2\, \Tr\, (g^\dagger t_A g t_3)
+\sin\beta_0~ 2\, \Tr\, (g^\dagger t_A gt_1)\ .\label{s7}
\eeqar
Thus all functions of these can be built up from the
$SO(3)$ elements $R_{AB}= 2 \Tr (g^\dagger t_A gt_B)$.
(Actually we need $B=1,3$, but $B=2$ is automatically
given by the cross product.)
Thus, in the large $l$ limit, the operators obeying
the further condition (\ref{s4}),
tend to the expected mode functions for the group
manifold of $SO(3)$ which is $S^3 /{\bf Z}_2$.
We have thus obtained a fuzzy version of
$S^3/{\bf Z}_2$ or ${\bf RP}^3$.
The condition we have imposed, namely (\ref{s4}), is
also very natural, once we realize that $(T-T')^2$
is the matrix analog of the Laplacian, and
mode functions can be obtained as eigenfunctions
of the Laplacian.

A similar construction is possible for fuzzy $S^4$, utilizing the fact  
that ${\bf CP}^3$
is an $S^2$-bundle over $S^4$. This has been shown in a recent paper by  
Abe
\cite{abe}.

\setcounter{equation}{0}
\section{Brane solutions in M(atrix) theory}

The idea of M-theory was formulated by Witten who showed that
the five superstring theories in ten dimensions could be considered as
special cases of a single theory, the M-theory \cite{witten}.
Witten also showed that eleven-dimensional supergravity is
another limit of M-theory, corresponding to compactification
on a circle.
Shortly afterwards, BFSS proposed a matrix model as a version of  
M-theory
in the lightcone formulation; as the dimension $N$
of the matrices becomes large, the model is supposed to describe
M-theory in the large lightcone momentum limit \cite{BFSS}.
Another matrix model, which applies to the type IIB case, as
opposed to the type IIA which is described by the BFSS model, has been  
given by
IKKT \cite{IKKT}. These matrix versions of M-theory are often referred  
to
as M(atrix)
models and have
been
rather intensively investigated over the last few years \cite{taylor}.
It is by now clear that M(atrix) theory does
capture many of the expected features of M-theory such as the
eleven-dimensional supergravity regime and the existence of
extended objects of appropriate dimensions.
Solutions to this theory are given by special matrices;
hence these theories generically have the possibility
of fuzzy spaces appearing as solutions.
These will correspond to brane solutions, with
smooth extended objects or branes emerging
in the large $N$-limit.
Finding such solutions is clearly of some interest.
The emergence of the two-brane or the standard membrane was analyzed
many years ago \cite{dewit}.
More recently, spherically symmetric membranes have been obtained  
\cite{kabat}.
As regards the five-brane, which is the other extended object of
interest, there has been no satisfactory construction or understanding
of the transverse brane where all five spatial dimensions are a subset
of the nine manifest dimensions of the matrix theory. The longitudinal
five-branes, called $L5$-branes, which have four manifest dimensions
and one along the compactified direction (either the eleventh dimension
or the lightlike circle) have been obtained. These include flat
branes \cite{banks} and stacks of $S^4\times S^1$-branes  
\cite{castelino}.
With the discussion of fuzzy spaces given in the previous
sections, we are now in a position to
analyze the construction of brane configurations. The presentation in  
this
section will closely follow
\cite{NR1}.

\subsection{The ansatz for a solution}

The action for matrix theory can be written as
\beq
{\mathcal S}= {{\Tr}} \left[ {{\dot X}_I^2\over 2R} +{R\over  
4}\,[X_I,X_J]^2
+\theta^T{\dot \theta}+i R\, \theta^T \Gamma_I [X_I,\theta]\right]
\label{(sol1)}
\eeq
where $I,J=1,...,9$ and $\theta$ is a $16$-component spinor
of $O(9)$ and $\Gamma_I$ are the appropriate gamma matrices.
The $\theta$'s represent the fermionic degrees of freedom
which are needed for the supersymmetry of the model.
$X_I$ are hermitian $(N\times N)$-matrices; they are elements
of the Lie algebra of $U(N)$ in the fundamental representation.
The theory is defined by this Lagrangian supplemented by the
Gauss law constraint
\beq
[X_I,{\dot X}_I] -[\theta, \theta^T] \approx 0 \ .\label{(sol2)}
\eeq
In the following we shall be concerned with bosonic
solutions and the $\theta$'s will be set to zero.
The relevant equations of motion are thus
\beq
{1\over R}\,{\ddot X}_I ~-~ R [X_J,[X_I,X_J]] =0\ .
\label{(sol3)}
\eeq
Even though the solutions we consider will be mostly fuzzy
${\bf CP}$'s, we will start with a more general framework,
analyzing general features of solutions for the model
(\ref{(sol1)}).
We shall look for solutions which carry some amount of symmetry.
In this case, we can formulate
simple ansatz in terms of the
group coset space $G/H$ where $H\subset G \subset U(N)$.
The ansatz we consider will have spacetime symmetries,
a spacetime transformation being compensated by an
$H$-transformation. Along the lines of the discussions in the previous
sections,
in the large $N$-limit, the matrices $X_I$
will go over to continuous brane-like solutions with the geometry
of $G/H$.

As before, let $t_A,~A=1,...,dimG$ denote a basis of the Lie algebra of
$G$. We split this set of generators into two groups, $t_\alpha,
~ \alpha =1,...,dimH$, which form the Lie algebra of $H$, and
$t_i,~i=1,..., (dimG-dimH)$, which form the complementary
set. Our ansatz will be to take $X_I$'s to lie along the
entire algebra of ${\underline G}$ or to be  linear combinations
of the $t_i$'s. In the latter case, in order to satisfy the equation of  
motion
(78), we shall then need the double commutator $[X_J,[X_I,X_J]]$
to be combinations of the $t_i$'s themselves. This is guaranteed
if $[t_i,t_j]\subset {\underline H}$, since the $t_i$'s themselves
transform as representations of $H$.
In this  case the commutation rules are of the form
\beqar
[t_\alpha ,t_\beta ]&=& if_{\alpha \beta \gamma } t_\gamma \  
,\nonumber\\[1mm]
{}[t_\alpha ,t_i] &=& i f_{\alpha ij} t_j \ ,\nonumber\\[1mm]
{}[t_i,t_j] &=& ic_{ij \alpha} t_\alpha \label{(sol4)}
\eeqar
and $G/H$ is a symmetric space. If $H=1$,
the $t_i$'s will belong to the full algebra ${\underline G}$.
In this case, $c_{ij}^\a$ of (\ref{(sol4)})
will be the structure constants of ${\underline G}$.
In such a case, even though the ansatz involves the full algebra
${\underline G}$, it can satisfy further algebraic constraints.
These additional constraints may have only
a smaller invariance group $H$; the solution will then again reduce to
the $G/H$-type.
These are the cases we analyze.

Since $X_I$ are elements of the Lie algebra of $U(N)$,
having chosen a $G$ and an $H$, we must consider the
embedding of $G$ in $U(N)$.
This is done as follows. We consider a value of $N$ which
corresponds to the dimension of a unitary irreducible
representation (UIR) of $G$. The embedding is then specified
by identifying the fundamental $N$-dimensional representation
of $U(N)$ with the $N$-dimensional UIR of $G$. Eventually we need
to consider the limit $N\rightarrow \infty$ as well.
Thus we must have an infinite sequence of UIR's of
$G$, so that we have a true fuzzy version
of $G/H$.
Generally, different choices of such sequences are possible,
corresponding to different ways of defining the
$N\rightarrow \infty$ limit.
Following previous sections, we will focus on
the symmetric rank $n$-tensors of $G$, of dimension,
say, $d(n)$. Thus we choose $N=d(n)$, defining the large
$N$-limit by $n\rightarrow \infty$.

The ansatz we take is of the form
\beq
X_i= r(t)\, {T_i\over N^a} \label{(sol5)}
\eeq
for a subset $i=1,...,p$ of the nine $X$'s. $T_i$ are generators
in ${\underline G}- {\underline H}$, in the symmetric rank $n$-tensor
representation of $G$. The eigenvalues for any of the $T_i$'s in the
$n$-tensor representation will range from $cn$ to $-cn$, where
$c$ is a constant. The eigenvalues thus become dense with a
finite range of the variation of the $X_i$'s  as $n\rightarrow \infty$
if $N^a \sim n$. In this case, as $n\rightarrow \infty$, the $X_i$'s
will tend to a smooth brane-like configuration. Our choice of the
index $a$ will be fixed by this requirement, viz., $N^a\sim n$.
(For fuzzy ${\bf CP}^k$, we defined $X$ by equation (\ref{45}).
they correspond to a particular choice of $a$. We will return to that  
choice
shortly.)
Notice also that this ansatz is consistent with the Gauss law  
(\ref{(sol2)}).
$r(t)$ represents the radius of the brane; it can vary with time and is  
the
only
collective coordinate in our ansatz.

The ansatz (\ref{(sol5)}) has a symmetry of the form
\beq
R_{ij} ~U X_j U^{-1} = X_i \label{(sol6)}
\eeq
where $R_{ij}$ is a spatial rotation of the $X_i$'s and $U$ is
an $H$-transformation for the $G/H$ case or more generally it can be in
$U(N)$. $R_{ij}$ is determined by
the choice of the $X_i$'s involved in (\ref{(sol5)}). Further, the  
ansatz
(\ref{(sol5)}) is to be
interpreted as being given in a specific gauge. A $U(N)$-transformation,
common to all the $X_i$'s, is a gauge transformation and does not
bring in new degrees of freedom. We may alternatively say that the  
meaning of
(\ref{(sol6)}) is that $X_j$ is invariant under rotations $R_{ij}$ up  
to a gauge
transformation.

For the ansatz (\ref{(sol5)}), the action simplifies to
\beq
{\mathcal S}= A_n \left[ {{\dot r}^2 \over 2RN^{2a}} ~-~
{{c_{ij\alpha}c_{ij\alpha}\over 4 N^{4a}}}\,Rr^4 \right]
\label{(sol7)}
\eeq
where $A_n$ is defined by ${\Tr} ( T_AT_B) =A_n \delta_{AB}$ and the
matrices and trace are in the $n$-tensor of $G$.
{}From its definition,
$A_n = d(n) C_2(n)/dimG$, where $C_2(n)$ is the quadratic Casimir
of the $n$-tensor representation, which goes like $n^2$ for large $n$.
Thus, with $N^a\sim n$, the
kinetic term in (\ref{(sol7)}) will always go like
$N/R~$ for large $n$.

We now turn to some specific cases. Consider first $G=SU(2)$.
In this case, $d(n)=n+1,~A_n= n(n+1)(n+2)/12$. The smooth brane
limit thus requires $a=1$ or $N^a \sim N \sim n$.  The kinetic
energy term in (\ref{(sol7)}) goes like $N/R$, while the potential term
goes like $R/N$. Thus both these terms would have a finite
limit if we take $N\rightarrow \infty,~R\rightarrow \infty$,
keeping $(N/R)$ fixed. In fact, this particular property holds
only for $G=SU(2)$ or products thereof, such as $G=O(4)$. For this
reason some of the branes which are realized as
cosets of products of the $SU(2)$ group can perhaps be regarded
as being transverse as their energy will not depend on $R$
in the large $R$ limit.
We can use this case to obtain a slightly
different description of the spherical membrane of \cite{kabat} as well  
as
  a "squashed" $S^2$ or ${\bf CP}^1$.
The round ${\bf CP}^1$ corresponds to the case
where the three generators of $SU(2)$ lie along three of the
nine $X_I's$.

As another example, consider $G=O(6)\sim SU(4)$. In this case,
$d(n)= (n+1)(n+2)(n+3)/6, A_n= (1/240)(n+4)!/(n-1)!$ and we need
$a= {1\over 3}$. The kinetic energy term goes like $N/R$
while the potential energy goes like $n R$. This corresponds to the
longitudinal five-brane with
$S^4$-geometry discussed in \cite{castelino}.
$SU(4)$ has an $O(5)$ subgroup under which the $15$-dimensional
adjoint representation of $SU(4)$ splits into the adjoint
of $O(5)$ and the $5$-dimensional vector representation.
The coset generators, corresponding to the $5$-dimensional
representation, may be represented by the $(4\times 4)$-gamma
matrices, $\gamma_\mu, ~\mu=1,...,5$.
The ansatz (5) thus takes the form
$X_\mu = r \gamma_\mu /N^{1\over 3}$. For the $N$-dimensional
representation, we use the symmetrized tensor product form given in
(\ref{s2}).  The sum of the squares of the $X_\mu$'s is
then proportional to the identity, thus giving effectively a
four-dimensional brane. This is interpreted as $n$ copies of
a longitudinal five-brane, one of the directions being along the
compactified eleventh or lightcone coordinate, of extent $R$,
an interpretation consistent with
the potential energy $\sim n R$.

\subsection{Fuzzy ${\bf CP}^2$ as a brane solution}

We now turn to an especially interesting case of a static solution
to the matrix theory, which corresponds to fuzzy ${\bf CP}^2$ with
$G= SU(3)$  obtained in \cite{NR1}. In this case, $d(n) = (n+1)(n+2)/2
~\sim n^2$
and $A_n= (1/48) (n+3)!/(n-1)!$. We choose $a=\half$. The kinetic
energy again goes like $N/R$ for large $N$ while the potential
energy goes like $R$. The potential energy is independent
of $n$ and thus, for $r$ independent of $t$, we have a single
static smooth five-brane wrapped around the compactified dimension
in the $n\rightarrow \infty$ limit. In other words, the tension
defined by the static energy per unit volume remains finite as
$n\rightarrow \infty$.
An appropriate choice of $H$ in this case is $H= U(2)\sim SU(2)\times
U(1)$. The world volume geometry is then
${\bf CP}^2\times S^1$, where $S^1$ corresponds to the compactified
eleventh dimension.
The immersion of this solution in ${\bf R}^9$ can be complicated
and will depend on the
details of the ansatz. As we shall see in the next section
the coset embedding will produce a singular surface in a  
nine-dimensional
Euclidean space, while when the eight generators of
$SU(3)$ are set parallel to eight of nine $X_I$'s we shall
obtain the standard ${\bf CP}^2$ embedded in an $S^7$ contained in  
${\bf R}^9$.

Notice also that since
the effective mass for the degree of freedom corresponding to $r$ goes  
like
$N/R$,
oscillations
in $r$ are suppressed as $N\rightarrow \infty$ with $R$ fixed; in this  
limit
this configuration becomes a static {\it solution}. We can fix $r$ to  
any value
and time-evolution does not change this.

Using the wave functions given earlier, it is straightforward to see
that, in matrix elements, we may use
\beqar
T_{+i} \equiv T_i &=& (n+3)\,{ \bz_i\over {(1+\bz \cdot z )}}\  
,\nonumber\\
T_{-i} \equiv T_i^\dagger &=&
(n+3)\, {z_i \over {(1+\bz \cdot z )}}\ ,\label{(sol11)}
\eeqar
\vspace{-4mm}
\beq
{}[T_{+i} , T_{-j}] \equiv h_{ij} = (n+3)\,{(\delta_{ij} -\bz_i z_j  
)\over {
(1+\bz \cdot z )}}\ .
\label{(sol11a)}
\eeq

The ansatz for the five-brane may now be stated as follows.
The simplest case to consider is the following.
We define the complex combinations $Z_i = {1\over
\sqrt{2}} (X_i +i X_{i+2})$, $i=1,2$.
The symmetric ansatz is then given by
\vspace{-2mm}
\beqar
Z_i &=& r(t)\, {T_i\over \sqrt{N}}\ ,~~~~~~~~~~i=1,2\,, \nonumber\\[1mm]
X_i&=&0\,,, ~~~~~~~~~~~~~~~~~~~i=5,...,9\ . \label{(sol14)}
\eeqar
In the large $n$-limit,
\vspace{-2mm}
\beq
Z_i\approx r \, {n\over \sqrt {N}}\,
{{\bz_i} \over {(1+\bz \cdot z )}} \approx  \sqrt{2}\, r \,{{\bz_i}  
\over
{(1+\bz \cdot z )}}
\label{sol14a}
\eeq
realizing a continuous map from ${\bf CP}^2$ to the space ${\bf R}^9$.
This map is not one-to-one; the region $\bz \cdot z <1$ and the region
$\bz \cdot z >1$ are mapped into the same spatial region
$\vert Z \vert < r\sqrt{2}$, corresponding to a somewhat squashed ${\bf  
CP}^2$.

The standard ${\bf CP}^2$ is obtained by
considering an ${\bf R}^8$-subspace
of ${\bf R}^9$, whose coordinates can be identified with the $SU(3)$  
generators
as in (\ref{(sol5)}), i.e., $X_A= r T_A/{\sqrt{N}},~A=1,2,...,8$, and  
$X_9=0$.
Specifically, using the expression for the
$SU(3)$ generators in terms of
the $z$'s,
this ansatz is given by (recall that in the large $N$ limit
$n\approx \sqrt{2N}$)
\begin{displaymath}
\hspace{1cm} ~X_i = {r\over\sqrt{2}}~ {{{\bar z} \sigma^i z}\over{1 +  
{\bar z}z}}\ ,
\end{displaymath}
\vspace{-5mm}
\beqar
X_4+iX_5&=& {r\over\sqrt{2}}  ~{{ 2 z_1}\over{1 + {\bar z}z}}\  
,\nonumber\\[1mm]
X_6 +iX_7&=& {r\over \sqrt{2}} ~ {{ 2z_2 }\over{1 + {\bar
z}z}}\ ,\label{(sol15a)}\\[1mm]
X_8&=&{{r}\over {\sqrt{6}}}~ {{2 - z{\bar z}}\over{1 +
{\bar z}z}}\ ,\nonumber
\eeqar
\vspace{-6mm}
\begin{displaymath}
\hspace{-0.8cm} X_9 =0
\end{displaymath}
where  $\sigma^i$, $i=1,2,3$ are Pauli matrices.
Notice that the singularity of the squashed ${\bf CP}^2$ is removed by  
this
ansatz since the regions $\bz \cdot z <1$ and $\bz \cdot z >1$
are mapped to different regions  of the nine-dimensional
space. In fact it
is easy
to see by a direct inspection that the map is actually one-to-one.
Furthermore it is easily seen that
$\Sigma_{A=1}^{8} X_A X_A = {2 r^2\over {3}}$. Thus
our surface is embedded in an $S^7$.
In fact, we can use  the above relations
and express
$X^1, X^2$ and $X^3$ in terms of $ X^4,.. .,  X^8$ as
\beq
X^a= {3\over\sqrt{2}}~{{ \bar\zeta \sigma^a\zeta }\over{r + \sqrt{6}  
X_8}}
\label{(sol15b)}
\eeq
where $\zeta$ is a two-component vector defined by $ \zeta^{1}= {1\over
\sqrt{2}}( X_4+iX_5 )$
and $ \zeta^{2}= {1\over \sqrt{2}}( X_6+iX_7 )$.

The coset structure is clearer directly in terms of the ansatz
for (\ref{(sol14)}), which is why we started with this squashed ${\bf  
CP}^2$.
The smooth configuration (\ref{(sol15a)}) may be regarded as a  
relaxation of
(\ref{(sol14)}) along some of the ${\bf R}^9$-directions.

The action for ansatz (\ref{(sol14)}), (\ref{(sol15a)}) becomes
\beqar
{\mathcal S}&=& {(n+3)!\over {(n-1)!}}\left[ {{\dot r}^2\over 12 NR}  
~-~ {Rr^4
\over 8N^2}
\right]\nonumber\\[1mm]
&\approx& \left[ {N\over R} {{\dot r}^2\over 3} ~-~ {Rr^4 \over  
2}\right]
\left( 1+{\mathcal O}({1/n} )\right)\ .\label{(sol17)}
\eeqar
In terms of the world volume coordinates, we can also write
\beqar
{\mathcal S} &\approx&\!\! \int\!\!  {2~d^4z \over {\pi^2 (1+\bz \cdot  
z )^3}}
\left[ {n^4\over 2 NR}\, {\dot r}^2 \,{\bz \cdot z \over {(1+\bz \cdot  
z )^2}}
- Rr^4 {n^4\over 4N^2}\, {2-2\bz \cdot z +(\bz \cdot z )^2 \over
{(1+\bz \cdot
z )^2}}\right]\nonumber\\
\label{(sol18a)}\\
&\approx& \!\! \int\!\!  {2~d^4z \over {\pi^2 (1+\bz \cdot z )^3}}  
\left[ {N\over R}\,{{\dot r}^2
\over 3}
- {Rr^4\over 2}\right]\,. \label{(sol18b)}
\eeqar
Expression (\ref{(sol18a)}) applies to ansatz (\ref{(sol14)}),  
expression
(\ref{(sol18b)}) to ansatz
(\ref{(sol15a)}). The energy densities are uniformly distributed over  
the
world volume
for (\ref{(sol15a)}), but not for (\ref{(sol14)}).

The equation of motion for $r$ becomes
\beq
{N\over R}\, {\ddot f} +6 f^3 =0 \label{(sol19)}
\eeq
where $r= f/{\sqrt{R}}$. The effective mass for the degree of freedom
corresponding to $r$ is ${\textstyle {2\over 3}} (N/R)$. Thus in the
limit $N\rightarrow \infty$ with $R$ fixed, any solution with finite
energy would have to have
a constant $r$ or $f$. In this limit, we thus get a five-brane which is
a {\it static  solution} of
the matrix theory. Alternatively, if we consider $N,R\rightarrow  
\infty$ with
$(N/R)$ fixed, $f$ can have a finite value.
However, in this limit, the physical dimension of the
brane as given by $r$ would vanish.

Explicit solutions to (\ref{(sol19)})
may be written in terms of the  sine-lemniscate function as
\beq
f = A \sin {\rm lemn} \left( \sqrt{3R\over N} ~A (t-t_0)\right)\,.
\label{(sol20)}
\eeq

\subsection{M-theory properties of the fuzzy ${\bf CP}^2$}

There are some properties of the solution we found which are of interest
for matrix theory point of view.
We will briefly mention these for completeness.

First, regarding the spacetime properties of the solution, we note that  
the matrix
action (\ref{(sol1)}) is in terms of lightcone coordinates. As a  
result, the Hamiltonian
corresponding to this action gives
the lightcone component $T^{+-}$ of the energy-momentum tensor.
Other components of the energy-momentum
tensor
can be evaluated,
following the
general formula of \cite{kabat2, taylor}; they will act as a source for
gravitons, again
along
the general lines of \cite{kabat2, taylor}. The current ${\mathcal
T}^{IJK}$ which is the
source for the antisymmetric tensor field of eleven-dimensional
supergravity is another
quantity of interest. This vanishes for the spherically symmetric
configurations
considered in \cite{castelino}, since there are no invariant
$O(9)$-tensors of the appropriate rank and symmetry. However, for the  
${\bf
CP}^2$
geometry, there is the K\"ahler form and the possibility that
${\mathcal T}^{-ij}$ can be proportional to the K\"ahler form has to be
checked explicitly.
The kinetic terms of ${\mathcal T}^{-ij}$  which depend on ${\dot r}$
are easily seen to vanish for the solution (\ref{(sol15a)}), essentially
because of
the symmetric nature of the ansatz. For solution (\ref{(sol14)}), we  
find, by
direct evaluation,
\beqar
{\mathcal T}^{-1 {\bar 1}} &=&{\mathcal T}^{-2{\bar 2}} = -{i\over
60}\,{n^5\over N^2}\,
{\dot r}^2 r^2~+...\ ,\nonumber\\[1mm]
{\mathcal T}^{-1{\bar 2}}&=& {\mathcal T}^{-2{\bar 1}}=0\ .
\label{(sol23)}
\eeqar
Naively, this diverges as $n\rightarrow \infty$. However, as we have
noticed before,
in this limit, the solution becomes static, ${\dot r}=0$ and hence this
vanishes.
This holds for other components of ${\mathcal T}^{IJK}$ as well.
The nonkinetic terms in ${\mathcal T}^{IJK}$ are of the form
$R^2 r^6 ({n^5/N^3})$ and also vanish as $n\rightarrow \infty$.
Thus the source for the antisymmetric tensor field is zero in the
$n ~({\rm or}~ N)\rightarrow \infty$ limit.

Another important spacetime property has to do with supersymmetry.
The Lagrangian of the matrix theory is invariant under
supersymmetry transformations. The supersymmetry
variation of $\theta$ is given by
\beq
\delta \theta \equiv K\epsilon +\rho ={{1}\over{2}} \left[{\dot  
X}_I\Gamma_I
+[X_I,X_J]\Gamma_{IJ}
\right]\epsilon + \rho \label{(sol24)}
\eeq
where  $\epsilon$ and $\rho$ are $16$-component spinors
of $O(9)$.

Since $\delta \theta$ has $n$-dependent terms, the question of
supersymmetry is best understood by considering fermionic
collective coordinates. These are introduced by using the
supersymmetry variation (\ref{(sol24)}) with the parameters
$\epsilon,~\rho$ taken to be time-dependent. Upon substitution in
the Lagrangian, the term $\Tr [ \theta^T {\dot \theta}]$ generates
the symplectic structure for $(\epsilon , \rho )$. We can then
construct the supersymmetry generators for fluctuations around our
solution. If the starting configuration is supersymmetric, there
will be zero modes in the symplectic form so constructed and we
will have only a smaller number of fermionic parameters appearing
in $\Tr [\theta^T {\dot \theta}]$. Now, in the large $n$-limit, we
have $\Tr [\theta^T {\dot \theta}]\sim n^2 [ \epsilon^T K^TK
{\dot \epsilon} +\rho^T {\dot \rho} ]$ which goes to zero as
$n\rightarrow \infty$ if $\delta \theta \sim n^{-1-\eta},~\eta
> 0$. Thus if $\delta \theta$ vanishes faster than $1/n$, we can
conclude that the starting bosonic configuration is
supersymmetric.

Consider the squashed ${\bf CP}^2$
first.
The finiteness of the kinetic energy in the large
$N$ limit requires that the leading term in $r$ must be a constant,  
which is
how we obtained a static solution. The equation of motion then
shows that
${\dot r}$ must go like ${1\over N}\sim {1\over n^2}$.
In other words, we can write $r= r_{0} + {{1}\over {N}} r_{1} + ...$.
The ${\dot X}_I$-term of $\delta\theta$ thus vanishes to the order
required. The vanishing of $\delta \theta$
(or the Bogomol'nyi-Prasad-Sommerfield-like condition)
then becomes, to leading
order,
\beq
-{{8r^2_0}\over {N}}(\lambda_a L_a + \sqrt {3}\lambda_8 R_1)\epsilon  
+\rho=0
\label{(sol25)}
\eeq
where the set $\lambda_a,\lambda_8$, $a=1,2,3$ generate an $SU(2)\times
U(1)$ subgroup
of $SU(3)$ while the operators  $ L_a, R_1$ generate an $SU_L(2)\times  
U_R(1)$
subgroup of $O(4)\sim SU_L(2)\times SU_R(2)$.
In terms of $(16\times 16)~\Gamma$-matrices they are given by
\beqar
L_1&=&{{i}\over{4}}\, (\Gamma_1\Gamma_3 + \Gamma_4\Gamma_2)\  
,\nonumber\\[1mm]
L_2&=&{{i}\over{4}}\, (\Gamma_1\Gamma_2 + \Gamma_3\Gamma_4)\  
,\label{(sol26)}\\[1mm]
L_3&=&{{i}\over{4}}\, (\Gamma_2\Gamma_3 + \Gamma_1\Gamma_4)\ ,\nonumber
\eeqar
\vspace{-5mm}
\beq
R_1={{i}\over{4}}\, (\Gamma_1\Gamma_3 - \Gamma_4\Gamma_2)\  
.\label{(sol26a)}
\eeq
Notice that the $\epsilon$-term is of order $1/n$ since
$\lambda_a,~\lambda_8$ have eigenvalues of order $n$ and in the large  
$n$ limit
$N\approx \half n^2$; the $\rho$-term
is of order one. Thus condition (\ref{(sol25)}) is required for
supersymmetry as
explained above.
Further, in our problem
the $O(9)$ group is broken to $O(4)\times O(5)$.
With respect to this breaking, the
$16$-component spinor of $O(9)$ decomposes according to
${\underline {16}}= ((1,2),4) + ((2,1), 4)$, where $4$ denotes the  
spinor
of $O(5)$. In terms of the
$SU_L(2)\times U_R(1)$ subgroup generated by the $ L_a$ and $R_1$ we
have
$(1,2) = 1_1+1_{-1}$ and $(2,1)= 2_0$ where the subscripts denote the
$U(1)$ charges.
Clearly we have no singlets under $SU_L(2)\times U_R(1)$;  
$SU_L(2)$-singlets
necessarily carry $U(1)$ charges. Therefore the operator
$(\lambda_a L_a + \sqrt {3}\lambda_8 R_1)$ can neither annihilate  
$\epsilon$
nor can it be a multiple of the
unit operator in the $SU(3)$ space. Hence a nontrivial
$\epsilon$-supersymmetry cannot be compensated by a
$\rho$-transformation. Thus we have no supersymmetry.

The supersymmetry variation produces a $\theta$ of the form
$(\lambda_a L_a + \sqrt {3}\lambda_8 R_1)\epsilon$, where we set
$\rho =0$ for the moment. The contribution of this $\theta$ to the
Hamiltonian via the term $\Tr (\theta^T [X_i,\theta ])$ is zero
due to the orthogonality of the $SU(N)$ generators. In other
words, the configurations $(X_i,0)$ and $(X_i,\delta \theta )$
have the same energy. This gives a supersymmetric set of degenerate
configurations, or supermultiplets upon quantization. The starting
bosonic configuration is not supersymmetric but is part of a set of
degenerate configurations related by supersymmetry.

Consider now the solution (\ref{(sol15a)}). In this case also, a
$\rho$-transformation
cannot compensate for an $\epsilon$-transformation and the condition for
supersymmetry becomes $f_{IJK}\Gamma_I\Gamma_J \epsilon =0$, where
$f_{IJK}$ are the structure constants of $SU(3)$.  
$L_K=f_{IJK}\Gamma_I\Gamma_J$
obey the commutation rules for $SU(3)$,
and, indeed, this defines an
$SU(3)$ subgroup of $O(8)$. The spinors of $O(8)$ do not contain  
singlets
under this $SU(3)$ and hence there is again no supersymmetry.

There is, perhaps, no surprise in this lack of supersymmetry, since  
${\bf
CP}^2$
does not admit a spin structure; nevertheless, it is interesting to see  
how
it works out at the matrix level.

\subsection{Other Solutions}

So far we have focused mainly on fuzzy ${\bf CP}^2$
as a solution to M(atrix) theory.
There are other interesting configurations possible, some of which we  
have
already
mentioned. One could consider ${\bf CP}^k$ in general; in this case,
the required exponent $a$ is given by $1/n$
and the potential energy goes like $n^{k-2} R r^4$.
Thus it is only for
$k=2$ that the potential energy becomes independent of $n$.

The required condition on the double
commutators is
satisfied if the $X_I$ span the entire Lie algebra of $G$ as well. The
equations
of motion for all these
cases are generically of the form
\beq
\left({N\over R}\right)^{2a} {\ddot f} +C_2 (adj) f^3 =0 \label{(sol29)}
\eeq
where $C_2 (adj)$ represents the quadratic Casimir of $G$ in the
adjoint representation and
$r= f R^{1-a}$. Since there are only nine $X_I$, if $G$ is not a product
group, its dimension
for this type of solutions cannot exceed $9$.
The case of $G=SU(2)$ reproduces the spherical
membrane \cite{kabat}. In this case $a=1$, and, as noted before, both
kinetic and potential energies have
well-defined limits as $N,~R\rightarrow \infty$ with their ratio
fixed.

In the large $n$ limit, this solution has the form,
\beqar
Z &=&{X_1+iX_2 \over \sqrt{2}}
= r(t)\, {(n+2)\over (n+1)}\, { \bz\over {(1+\bz \cdot z )}}\approx
r(t)~ { \bz\over {(1+\bz \cdot z )}}\ ,\nonumber\\[1mm]
{\zeta} &=&X_3= \half\, r(t)\, {(n+2)\over (n+1)}\, {{(1- \bz\cdot
z)}\over {(1+\bz \cdot z )}}\approx \half \,r(t) ~{{(1- \bz\cdot
z)}\over {(1+\bz \cdot z )}}\ . \label{(sol30)}
\eeqar
Clearly we
have a two-sphere defined by
\beq Z\bZ +  \zeta^2 \approx {1\over
4}\,r(t)^2 \ .\label{(sol31)}
\eeq
The radius of the sphere remains
finite as $n\rightarrow \infty$. Even though the ansatz has the
full $SU(2)$-symmetry, there is a further algebraic constraint,
viz., (\ref{(sol31)}), and this reduces the space of free
parameters to $SU(2)/U(1)$.

Another interesting case which was briefly mentioned is that of
$S^4$-geometry which is related to the coset $O(6)/O(5)$, with a further
algebraic
condition which reduces the dimension to four \cite{castelino}.
There are other cases which can be considered along
these lines; for example, for $SU(2)\times SU(2)$, we can set
six of the $X_i$'s
proportional to the generators and the energies depend only on
$N/R$ as $N, R \rightarrow \infty$. This can be embedded in $U(N)$ in
a block-diagonal way by choosing the representation $(N_1,1)+(1,N_2)$
with $N=N_1+N_2$. Presumably this can give two copies of the two-brane
in some involved geometrical arrangement in ${\bf R}^9$.

\setcounter{equation}{0}
\section{Fuzzy spaces and the quantum Hall effect}

There is an interesting connection between the quantum Hall effect and
fuzzy spaces
which we shall briefly discuss now.

\subsection{The Landau problem and ${\mathcal H}_N$}

In the classic Landau problem of a charged particle in a magnetic
field ${\vec B}$, one has a  number of equally spaced Landau
levels. Other than the translational degree of freedom along the
magnetic field, the dynamics is confined to a two-dimensional
plane transverse to ${\vec B}$. In many physical situations, based
on energy considerations, the dynamics is often confined to one
Landau level, say the lowest. In this case, the observables are
hermitian operators on this subspace of the Hilbert space and are
obtained by projecting the full operators to this subspace. The
operators representing coordinates, for example, when projected to
the lowest Landau level (or any other level), are no longer
mutually commuting. The dynamics restricted to the lowest Landau
level is thus dynamics on a noncommutative two-plane. This has
been known for a long time. One can generalize such considerations
to a two-sphere, for example. (The Landau problem on the
two-sphere was considered by Haldane \cite{haldane}; we follow
\cite{KN1}.) One can have a uniform magnetic field on the
two-sphere which is normal to it; this would be the radial field
of a magnetic monopole sitting at the origin if we think of the
two-sphere as embedded in the usual way in ${\bf R}^3$.
(The potential for a uniform magnetic field is given by
(\ref{1.1}).) Since $S^2
= SU(2)/U(1)$, we may think of the wave functions for a particle
on $S^2$ as functions of $g \in SU(2)$ with the condition that
they are invariant under $g \rightarrow g h$, $h \in U(1)$, so
that they actually project down to $S^2$. We may take the $U(1)$
direction to be along the $T_3$ direction in the $SU(2)$ algebra.
Since a basis of functions for $SU(2)$ is given by the Wigner
${\mathcal D}$-functions, a basis for functions on $S^2$ is given
by the $SU(2)$ Wigner functions ${\D}^{(j)}_{mk} (g)$, with
trivial right action of $U(1)$, in other words, the
$U_R(1)$-charge, $k =0$. In this language, derivatives on $S^2$
can be identified as $SU(2)$ right rotations on $g$ (denoted by
$SU_R(2)$) satisfying an $SU(2)$ algebra
\beq \bigl[ R_{+},~R_{-}
\bigr] = 2~ R_{3}\label{qhe1}
\eeq
where $R_{\pm}= R_1 \pm i R_2$.
$R_{\pm}$ are dimensionless quantities. The standard covariant
derivatives, with the correct dimensions, are
\beq D_{\pm} =
i~{R_{\pm} \over r}\label{qhe2}
\eeq
where $r$ is the radius of
the sphere. In the presence of the magnetic monopole, the
commutator of the covariant derivatives is related to the magnetic
field, in other words, we need $[D_+ ,D_-] = -2B$. With the
identification (\ref{qhe2}), and the commutation rule
(\ref{qhe1}), we see that this fixes $R_{3}$ to be half the
monopole number $n$, with $n =2Br^2$. Therefore the wave functions
on $S^2$ with the magnetic field background are of the form
${\D}^{(j)}_{m,{n \over 2}} (g)$. The Dirac quantization rule is
seen, from this point of view, as related to the quantization of
angular momentum, as first noted by Saha \cite{saha}. For a
detailed description of the formalism presented here and analysis
of fields of various spin on $S^2$ on the monopole background see
\cite{Randjbar-Daemi:1982hi}.

We can now write down the one-particle Hamiltonian
\beqar
H &=& - {1 \over {4\mu}} \bigl( D_{+} D_{-} + D_{-} D_{+} \bigr)  
\nonumber \\
&=& {1 \over {2\mu r^2}}\, \Bigl( \sum_{A=1}^3 R_{A}^2 - R_3^2 \Bigr)
\label{qhe3}
\eeqar
where $\mu$ is the particle mass.
For the eigenvalue $\half n$ to occur as one of the possible values for
$R_3$, so that we can form ${\D}^{(j)}_{m,{n \over 2}} (g)$,
we need $j=\half n + q$, $q=0,1,..$.
Since $R^2=j(j+1)$,
the energy eigenvalues are
\beqar
E_q & = & {1 \over {2\mu r^2 }} \left[ (\half n + q) (\half n + q +1) -  
{n^2
\over 4} \right] \nonumber \\[1mm]
& = & {B \over {2\mu}}\, (2 q +1) + {{q(q+1)} \over {2\mu r^2}}\  
.\label{qhe4}
\eeqar
The integer $q$ plays the role of the Landau level index. The lowest
Landau
level ( $q =0$),or the ground state, has energy
$B/2\mu$, and the
states
$q > 0$ are separated by a finite energy gap. The degeneracy of
the
$q$-th Landau level
is $2j+1=n+1+2q$.
(Notice that, in the limit
$r
\rightarrow \infty$, the planar image of the sphere under
the
stereographic map becomes flat and so this corresponds to the
standard
planar Landau problem. We see that, as $r\rightarrow
\infty$,
(\ref{qhe4})
reproduces the known planar result for the energy
eigenvalues and the degeneracy.

In the limit of large magnetic fields, the separation of the
levels is large, and it is meaningful to restrict dynamics to one
level, say the lowest, if the available excitation energies are
small compared to $B/2\mu$. In this case, $j =\half n$, $R_3
=\half n$, so that we have the highest weight state for the right
action of $SU(2)$. The condition for the lowest Landau level is
$R_+ \Psi =0$ and this level has degeneracy $n+1$.

We now see the connection to fuzzy $S^2$. The Hilbert space of the
lowest Landau level corresponds exactly to the symmetric rank $n$
representation of $SU(2)$. The condition $R_+ \Psi =0$, which was
used as the condition restricting the wave functions to depend on
only half of the phase space coordinates in the quantization
procedure outlined in section 2, is obtained for the Landau
problem as well, but as a condition choosing the lowest Landau
level. The Hilbert subspace spanned by $\D^{({n\over
2})}_{m,{n\over 2}}$ is the same and hence, all observables for
the lowest Landau level correspond to the observables of the fuzzy
$S^2$.

This correspondence can be extended to the Landau problem
on other
spaces. For all ${\bf CP}^k$ with a $U(1)$ background field,
we have an
exact correspondence between the lowest Landau level
and fuzzy ${\bf
CP}^k$. The background field specifies the
choice of the eigenvalues $R_i$
in the Wigner ${\mathcal D}$-functions;
the lowest Landau level condition
becomes the polarization condition
for the wave functions \cite{KN1}.
The wave functions for the lowest Landau level are
exactly those given in (\ref{1.18}); they are characterized by the  
integer
$n$, which gives the rank of the symmetric $SU(k+1)$ representation
and corresponds to a uniform magnetic field along the direction
$t_{k^2+2k}$, as seen from (\ref{1.15}).

An
especially interesting case is that of
${\bf CP}^3$. Because this is an
$S^2$ bundle over
$S^4$, the Landau problem on ${\bf CP}^3$ is equivalent
to
a similar problem on $S^4$ with an $SU(2)$ background field
\cite{HZ}.
We will come back to this briefly.

\subsection{A quantum Hall
droplet and the edge excitations}

In discussing the physics of the
quantum
Hall effect, we need to go beyond just the construction of the
states.
Typically one has a number of states occupied by electrons,
which
are fermions, and so there is no double occupancy for any state.
Generally
these electrons cluster into a droplet.
Dynamically this is due to an
additional potential ${\hat V}$; electrons tend to
localize near the
minimum of the potential.  The excitations of this
droplet are of interest
in quantum Hall systems. Since there cannot be
double occupancy and there
is conservation of the number of electrons,
the excitations are
deformations of the droplet which preserve the total volume of
occupied
states. In the large $n$ limit, these are surface deformations of an
almost
continuous droplet; they are hence called the edge excitations.

We can
specify the droplet by a diagonal density matrix
${\hat \rho}_0$ which is
equal to $1$ for occupied states and zero for unoccupied states.
The
dynamical modes are then fluctuations which keep the number of
occupied
states, or the rank of ${\hat \rho}_0$, fixed. They are thus given by
a
unitary transformation of ${\hat \rho}_0$, ${\hat \rho}_0 \rightarrow  
{\hat
U} {\hat \rho}_0 {\hat U}^\dagger$.
One can write an action for these modes
as
\beq
S= \int dt\left[ i \Tr\, ({\hat \rho}_0 { \hat U}^\dagger \del_t
{\hat U})
- \Tr \,({\hat \rho}_0 {\hat U}^\dagger {\hat H} {\hat U})
\right]
\label{qhe5}
\eeq
where ${\hat H}$ is the Hamiltonian. Since we are
in the lowest Landau level
of fixed energy, we can take the Hamiltonian to
be just the potential
${\hat V}$.
Variation of ${\hat U}$ leads to the
extremization condition for
$S$ as
\beq
i \,{\del {\hat \rho}\over \del t}
=
[ {\hat H} , {\hat \rho}]
\label{qhe6}
\eeq
which is the expected evolution
equation for the density
matrix.
In the large $n$ limit,
we can simplify
this action by writing ${\hat U} = \exp (i {\hat \Phi})$,
and replacing
operators by their symbols,
matrix products by $*$-products and the trace
by
${\bf CP}^k$-integration, as discussed in section 2.
We also have to
write ${\hat U} \rightarrow
1 + i \Phi - {1\over 2!} \Phi * \Phi +\cdots$,
where $\Phi$ is the symbol
for ${\hat \Phi}$. We will consider a droplet
with $M$ occupied states,
with $M$ very large.
The large $n$ limit of
(\ref{qhe5}) can then be obtained,
for a simple spherical droplet, as
\cite{KN1}
\beq
S_{{\bf CP}^k}\approx -{1\over 4\pi^k} M^{k-1}\int
d\Omega_{S^{2k-1}}
  \left[ {\del\Phi
\over \del t}\, (\L\Phi ) +
\omega
\left( \L\Phi \right)^2 \right]
\label{qhe7}
\eeq
where
$d\Omega_{S^{2k-1}}$ denotes the volume element on the sphere
$S^{2k-1}$,
which is the boundary of the droplet; the factor $M^{k-1}$ is as  
expected
for
a droplet of radius $\sim \sqrt{M}$.
The operator $\L$ is identified in
terms of the coordinates
$\bz ,z$ as
\beq
\L = i \left( z\cdot {\del \over
\del z} -
\bz\cdot {\del \over \del \bz}\right)\,.
\label{qhe8}
\eeq
Terms which vanish as
$n \rightarrow \infty $ have been dropped.
As an example,
we have taken a potential
\beq
{\hat V}= \sqrt{2k \over k+1}~\omega \left(
T_{k^2+2k} +
{nk\over \sqrt{2k(k+1)}}\right)
\label{qhe9}
\eeq
with
$\omega$ as a constant parameter.
(The form of the action is not sensitive
to the specifics of the potential;
more generally one has $\omega = {1\over
n} {\del V \over \del (\bz\cdot z)}$.)
The action (\ref{qhe7}) is a
generalization of a chiral Abelian Wess-Zumino-Witten
theory. (The
calculations leading to (\ref{qhe7}) are not complicated, after the
discussion
of the large $n$ limit in section 2; but they are still quite
involved and we refer the reader
to the original articles.)

One can also
consider non-Abelian background fields,
say, constant $SU(k)$ backgrounds
for ${\bf CP}^k$, since the latter is
$SU(k+1)/U(k)$. In this case, the
wave functions must obey
the conditions
\beqar
{\hat R}_a ~\Psi^J_{m,
\alpha} (g) &=&
(T^{{\tilde J}}_a)_{\alpha \beta} \Psi^J_{m, \beta} (g)\ ,
\nonumber\\[1mm]
{\hat R}_{k^2 +2k} ~\Psi^J_{m, \alpha} (g) &=& - {n k\over
\sqrt{2 k
(k+1)}}~\Psi^J_{m, \alpha} (g) \label{qhe10}
\eeqar
since there
is a background  $SU(k)$-field.
The wave functions must transform under
right
rotations as a  representation
of
$SU(k)$,
$(T^{{\tilde
J}}_a)_{\alpha \beta}$ being the representation matrices for the
generators
of
$SU(k)$ in the representation ${\tilde J}$.
$n$ is an integer
characterizing the Abelian part of the background field.
$\alpha ,\beta$
label states within the $SU(k)$ representation ${\tilde J}$
(which is
itself
contained in the representation $J$ of $SU(k+1)$). The index
$\alpha$ carried by the
wavefunctions (\ref{qhe10}) is basically the gauge
index. The wave functions are sections
of a
$U(k)$ bundle on ${\bf
CP}^k$.
The wave functions for the lowest Landau level are thus given
by
\beqar
\Psi^J_{m, \alpha} (g) &=& \sqrt{N} ~\la J, L \vert ~{\hat g}
~\vert J, ({\tilde
J},\alpha ), - n\ra\nonumber\\[1mm]
&=& \sqrt{N}~ {\mathcal
D}^J_{m; \alpha}(g)\ .\label{qhe11}
\eeqar

The symbol corresponding to an
operator ${\hat F}$ is now a matrix, defined as
\beq
F_{\alpha \beta}(g) =
\sum_{km}{\mathcal D}_{k,\alpha}(g) ~F_{km}
{\mathcal
D}^*_{m,\beta}(g)\ .
\label{qhe12}
\eeq
The simplification of the
action (\ref{qhe5}) will now involve a field
$G$ which is a unitary matrix,
an element of $U(dim {\tilde J})$.
The large $n$ limit can be calculated
as
in the Abelian case and gives the action \cite{KN2}
\beqar
{\mathcal
S}(G)&=& {1 \over {4 \pi^k}} M^{k-1} \int_{\del {\mathcal D}}
dt ~ \tr
\left[ \left( G^{\dagger} {\dot G} ~+~ \omega~G^{\dagger} {\mathcal L} G
\right)
G^{\dagger}{\mathcal L}G \right] \nonumber\\[1mm]
~&&~+ (-1)^{{k(k-1)}
\over 2}  {i\over 4\pi }\, {M^{k-1}\over
(k-1)!}
\int_{\mathcal D} dt  ~2\,
\tr \left[
  G^{\dagger} {\dot G} (G^{-1}D G)^2 \right]\wedge \left({{i
\Omega} \over \pi} \right)^{k-1}\!\! .
  \nonumber\\
\label{qhe13}
\eeqar
This is
a chiral, gauged Wess-Zumino-Witten (WZW) action generalized to
higher
dimensions.
Here the first term is
on the boundary $\del {\mathcal
D}$
of the droplet and it is precisely the gauged, non-Abelian analog of
(\ref{qhe7}). The operator ${\mathcal L}$ in
(\ref{qhe13}) is the gauged
version of (\ref{qhe8}),
\beq
{\mathcal{L}} = i \bigl( z^i D_i - \bz^i
D_{\bar i} \bigr)\ .
\label{qhe14}
\eeq
The gauge covariant derivative is
given by $D = \del + [
\A,~~ ]$, where
$\A$ is the
$SU(k)$ gauge potential,
given by $\A^a_i = 2i \Tr ( t^a g^{-1} dg)$.
(This potential corresponds to
the spin connection on ${\bf CP}^k$;
the corresponding Riemann curvature is
constant in the tangent frame basis.
The gauge field we have chosen is
proportional to this.)
The second term in (\ref{qhe13}), written as a
differential form, is a higher dimensional
Wess-Zumino term; it is an
integral over
the droplet ${\mathcal D}$ itself, with the radial variable
playing the role
of the extra dimension.
As expected, since we have an
$SU(k)$ background,
the action has an $SU(k)$  gauge symmetry.

We shall
discuss the fuzzy space point of view regarding the edge excitations
in
quantum Hall effect
shortly; before we do that, we shall briefly consider
Hall effect on spheres.

\subsection{Quantum Hall effect on spheres}

The
most interesting cases of quantum Hall effect on spheres pertain to
$S^4$
and $S^3$.

The edge excitations for the droplet on $S^4$ was recently
suggested
by Zhang and Hu as a model for higher spin gapless states,
including the
graviton \cite{HZ}. (In fact, this is what started many
investigations into
higher dimensional quantum Hall effect \cite{everyone,
pol}.)
This is a very nice idea, although it has not yet worked out
as
hoped for. The action for this case
can be obtained from (\ref{qhe7})
by
utilizing the fact
that ${\bf CP}^3$ is an $S^2$-bundle over $S^4$.
We can
describe ${\bf CP}^3$ by the four complex coordinates
$Z_\alpha$,
$\alpha=1,...,4$, with the identification
$Z_\alpha \sim \lambda Z_\alpha$
where $\lambda$ is any complex number
except zero, $\lambda \in {\bf C}-\{
0 \}$. Explicitly, we may write
$Z_\alpha$ as $\sqrt{{\bar Z}\cdot
Z}~u_\alpha = \sqrt{{\bar Z}\cdot Z}~g_{\alpha 4}$, but for the present
purpose,
it is more convenient to write it in terms of
two
two-component
spinors $w ,~\pi$ as
\beq
(Z_1, Z_2, Z_3, Z_4)= (w_1, w_2,
\pi_1, \pi_2)\ .
\label{qhe15}
\eeq
Coordinates $x_\mu$ on $S^4$ are then
defined by
\beq
w = (x_4 -i \sigma \cdot x ) ~\pi\ .
\label{qhe16}
\eeq
The
scale invariance $Z\sim \lambda Z$ can be realized
as the scale invariance
$\pi \sim \lambda \pi$;
the $\pi$'s then describe a ${\bf CP}^1 =S^2$. This
will be
the fiber space. The coordinates $x_\mu$ are the
standard
stereographic coordinates for $S^4$; one can in fact
write
\beq
y_0 = {1-x^2 \over 1+x^2}\ ,\qquad\quad y_\mu = {2x_\mu \over
1+x^2}
\label{qhw17}
\eeq
to realize the $S^4$ as embedded in ${\bf
R}^5$.
The definition of $x_\mu$ in terms of $w$ may be
solved
as
\beqar
x_4 &=& {1\over 2}\, {{\bar \pi}w +{\bar w}\pi \over
{\bar
\pi}\pi}\ ,\nonumber\\[1mm]
x_i&=& {i\over 2}\, {{\bar \pi} \sigma_i w - {\bar
w}
\sigma_i \pi \over {\bar \pi} \pi}\ .
\label{qhe18}
\eeqar
There is a
natural subgroup, $SU_L(2) \times SU_R(2)$, of $SU(4)$,
with $\pi$
transforming as the fundamental representation
of $SU_L(2)$ and $w$
transforming as the fundamental representation
of $SU_R(2)$.

The K\"ahler
two-form on ${\bf CP}^3$ is given, as in (\ref{1.3}), by
\beq
\Omega = - i
\left[ {d{\bar Z}\cdot dZ \over {\bar Z}\cdot Z}
- {d {\bar Z}\cdot Z
~{\bar Z}\cdot dZ \over ({\bar Z}\cdot Z)^2}
\right]\,. \label{qhe19}
\eeq
This is invariant under $Z\rightarrow \lambda Z$,
and ${\bar Z}\rightarrow
\lambda {\bar Z}$.
We can reduce this using (\ref{qhe16}), (\ref{qhe18})
to get
\beqar
\Omega_{{\bf CP}^3} &=& \Omega_{{\bf CP}^1} -
i~F\ ,\nonumber\\[1mm]
F&=& dA +A ~A\ ,\nonumber\\[1mm]
A&=& i\, {N^a \eta^a _{\mu\nu} x^\mu
dx^\nu \over (1+x^2)}
\label{qhe20}
\eeqar
where
\beqar
\eta^a_{\mu\nu} &=&
\ep_{a\mu\nu 4} + \delta_{a\mu} \delta_{4\nu}
- \delta_{a\nu} \delta_{4\mu}
\nonumber\\[1mm]
N^a &=& {\bar \pi} \sigma^a \pi /{\bar
\pi}\pi\label{qhe22}
\eeqar
$\eta^a_{\mu\nu}$
is the 't Hooft tensor
and
$N^a$ is a unit
three-vector, which may be taken as parametrizing the
fiber
${\bf CP}^1 \sim S^2$.
The field $F$ is the instanton field. We see
that we
can get an instanton background on $S^4$ by taking
a $U(1)$
background field on ${\bf CP}^3$ which is proportional to the
K\"ahler
form.

The action (\ref{qhe7}) may now be used with the separation of
variables
indicated;
it simplifies to
\beq
S = - {M\over 4\pi^2} ~n\int
d\mu_{{\bf CP}^1}~
\int d\Omega_3 \left[ {\del\Phi \over \del t} (\L\Phi )
+
\omega (\L \Phi )^2 \right]\label{qhe23}
\eeq
where $\L\Phi = 2x^\nu
K^{\mu\nu} \del_\mu \Phi$. $\Phi$'s are to
be expanded
in terms of
harmonics on ${\bf CP}^1$ which correspond to
the representations of
$SU_L(2)$ in $SU_L(2)\times SU_R(2)$. (This is the subgroup
corresponding
to the instanton gauge group.)
Since $\Phi$ is a function on
${\bf CP}^1$,
we must have invariance under the scaling $\pi \rightarrow
\lambda \pi$.
The mode expansion for $\Phi$ is thus given by \cite{KN1}
\beqar
\Phi &=&
\sum_{l\geq m} ~
C^{(\dot A)_m (\dot B)_l (C)_{l-m+k} (D)_k} ~f_{(\dot A)_m
(\dot B)_l (C)_{l-m+k}
(D)_k}\nonumber\\
&&\hskip .3in +\sum_{l < m}
~
{\tilde C}^{(\dot A)_m (\dot B)_l (C)_{k} (D)_{m-l+k}}~{\tilde f}_{(\dot
A)_m (\dot B)_l
(C)_{k} (D)_{m-l+k}}\label{qhe24}
\eeqar
where the mode functions have the form
\beqar
f_{(\dot A)_m (\dot B)_l
(C)_{l-m+k}
(D)_k}&=&
{1\over ({\bar \pi}\cdot \pi )^{l+k}}~{\tilde
w}_{\dot A_1}
\cdots  {\tilde w}_{\dot A_m}
w_{\dot B_1} \cdots w_{\dot
B_l}\nonumber\\
&&\hskip .7in \times  {\tilde \pi}_{C_1} \cdots {\tilde
\pi}_{C_{l-m+k}}
\pi_{D_1} \cdots \pi_{D_k}\ , \nonumber\\[1mm]
{\tilde f}_{(\dot
A)_m (\dot B)_l
(C)_{k} (D)_{m-l+k}}
  &=& {1\over ({\bar \pi}\cdot \pi
)^{m+k}}~{\tilde w}_{\dot A_1}
\cdots  {\tilde w}_{\dot A_m}
w_{\dot B_1}
\cdots w_{\dot B_l}\nonumber\\
&&\hskip .7in \times {\tilde \pi}_{C_1}
\cdots {\tilde \pi}_{C_k} \pi_{D_1}
\cdots
\pi_{D_{m-l+k}}
\label{qhe25}
\eeqar
where $({\dot A})_m = {\dot
A}_1 \cdots {\dot A}_{m}$, $(C)_k = C_1 \cdots C_k$ and similarly
for the
other indices. Each function $f$ ($\tilde{f})$ transforms as an
irreducible
representation of $SU_L(2) \times SU_R(2)$, with
the $j$-values $\half
\vert l-m\vert + k$ and $\half (l+m)$ respectively.
The action
(\ref{qhe23}) and the mode expansion (\ref{qhe24})
show clearly the
emergence of the higher spin gapless
modes. In particular, it is possible
to obtain massless
spin-$2$ excitations.
The difficulty, however, is that
there are many other modes which do not all
combine into a relativistically
invariant theory \cite{pol, KN1}. Perhaps, some clever
projection, such as
the GSO projection in string theory, may be possible.

One can formulate
quantum Hall
effect on other spheres as well. For $S^3$, for example, we
can use
the fact that $S^3 \sim SU(2) \times SU(2) /[SU(2)]$; this shows
that it is
possible to have a constant $SU(2)$ gauge field on $S^3$
(which
will be proportional to the Riemann curvature
of $S^3$).
Taking this gauge
field one can obtain Landau level
states and an edge action \cite{NR2}.
If
we denote the generators
of the Lie algebra of the two $SU(2)$'s by
$L_a$
and $R_a$, we can take the derivatives on
$S^3$ to be proportional to $L_a
- R_a$, with the
$SU(2)$ being divided out defined by $J_a = L_a+R_a$.
The
Landau levels will correspond to
the Wigner ${\mathcal D}$-functions of
$SU(2) \times SU(2)$, with
the representation under
$J_a$'s specifying the
background field.

In the case of the ${\bf CP}^k$'s discussed earlier, the lowest
Landau level corresponds to the Hilbert space ${\mathcal H}_N$ of
the fuzzy version of the space. This suggests that one can utilize
the construction of Landau levels on $S^3$, and more generally on
other spheres, to get a definition of fuzzy spheres. Actually, for
the three-sphere, the lowest Landau level corresponds, not to a
fuzzy $S^3$, but a fuzzy $S^3/{\bf Z}_2$  \cite{NR2}; the
realization of $S^3 / {\bf Z}_2$ is essentially identical to our
discussion in section 5. Spheres of other dimensions can be
considered using the fact that $S^k = SO(k+1)/ SO(k)$; constant
fields which correspond to $SO(k)$ gauge fields are then possible
and one can carry out a similar analysis for the quantum Hall
effect.

\subsection{The fuzzy space-quantum Hall effect connection}

We
now return to the question of what quantum Hall effect has to do with  
fuzzy
spaces.

Fuzzy spaces are based on the trio $({\mathcal H}_N, Mat_N ,
\Delta_N)$.
The Hilbert space ${\mathcal H}_N$ is obtained by quantization
of the action
(\ref{1.10}); the wave functions are sections of
an
appropriate $U(1)$ bundle on the space $M$ whose fuzzy version we are
constructing.
$Mat_N$ is then the matrix algebra of linear transformations
of this Hilbert space.
The lowest Landau level for quantum Hall effect on a
compact manifold
$M$, as we mentioned before, defines a finite dimensional
Hilbert
space which is identical to ${\mathcal H}_N$. This is clear,
since,
  with  a background magnetic field, the wave functions are sections
of a $U(1)$ bundle
  on $M$.
Thus observables of the quantum Hall system
are
elements of $Mat_N$.

We can go further and ask how we may characterize
subspaces
of fuzzy spaces. A region, which is topologically a disk, may be
specified by
a projection operator.
We assign a  value $1$ to
the
projection operator for states inside the region and zero for
states
outside the region. Notice that this is precisely what the droplet
density
operator does.  The fluctuations of the projection operator
preserving its rank
are the analogs of volume-preserving transformations.
In the large
$n$ limit, they correspond to the field $\Phi$. We may thus
regard
${\hat U}$ as specifying the modes corresponding to different
embeddings of
a fuzzy disk in the full fuzzy space \cite{lizzi}. Clearly
this is of geometrical interest.
In fact, we can go a bit further with this
analogy.
The action for ${\hat U}$, namely (\ref{qhe5}), is the same
as the
action (\ref{1.10}), except that the group is now $U(N)$ and
the invariant
subgroup is chosen by the projection operator or density matrix.
The
quantization of this action will thus lead to another fuzzy space,
which
will correspond to the
set of scalar fields (corresponding to $\Phi$) on
the boundary of
the chosen region.
(For these arguments, we may even set
${\hat V} =0$.)

The case of the non-Abelian background is presumably related to
vector bundles rather than functions on the fuzzy space, in a way
which is not yet completely clarified. The fact that this leads to
  an action of the WZW type in the large $n$ limit is also quite
interesting. From what we have said so far, it is clear that there
is a set of concepts linking fuzzy spaces and the quantum Hall
effect, with the possibility of a more fruitful interplay of
ideas.

\section*{Acknowledgments}
The work of DK and VPN was supported in part by the National
Science Foundation under grant numbers PHY-0140262 and
PHY-0244873.

\end{document}